\documentclass[twocolumn]{aastex63}

\received{XX}
\revised{YY}
\accepted{ZZ}

\submitjournal{ApJ}

\shorttitle{Chemical clocks in wide binaries}
\shortauthors{Espinoza-Rojas et al.}

\graphicspath{{./}{figures/}}

\usepackage{xcolor}
\begin{document}

\title{The consistency of chemical clocks among coeval stars} \footnote{Released on XX, YY, ZZ}

\correspondingauthor{Francisca Espinoza Rojas}
\email{fmespinoza@uc.cl}

\author{Francisca Espinoza-Rojas}
\affiliation{Instituto de Astrofísica, Pontificia Universidad Católica de Chile \\ Av. Vicuña Mackenna 4860, 782-0436 Macul, Santiago, Chile}

\author{Julio Chanamé}
\affiliation{Instituto de Astrofísica, Pontificia Universidad Católica de Chile \\ Av. Vicuña Mackenna 4860, 782-0436 Macul, Santiago, Chile}

\author{Paula Jofré}
\affiliation{Núcleo de Astronomía, Facultad de Ingeniería y Ciencias,  Universidad Diego Portales\\ Ejército 441, Santiago, Chile}

\author{Laia Casamiquela}
\affiliation{Laboratoire d’Astrophysique de Bordeaux, Univ. Bordeaux \\ CNRS, B18N, allée Geoffroy Saint-Hilaire, 33615 Pessac, France}

\begin{abstract}

The abundance ratios of some chemical species have been found to correlate with stellar age, leading to the possibility of using stellar atmospheric abundances as stellar age indicators. These chemical clocks have already been calibrated with solar twins and open clusters, but it remains to be seen whether they can be effective at identifying coeval stars in a population that spans a broad parameter space (i.e., the promise of chemical tagging). Since the components of wide binaries are known to be stars of common origins, they constitute ideal laboratories for testing the usefulness of chemical clocks for the age dating of field stars. Using a combination of our new measurements and literature data on wide binaries, we show for the first time that chemical clocks are even more consistent among the components of wide binaries than their individual abundances. Moreover, the special case of HIP 34426/HIP 34407 may indicate that chemical clocks are consistent for coeval stars even when those individual abundances are not. If the assumption that chemical clocks are reliable age indicators is correct, this would constitute first statistically significant evidence that the components of wide binaries are indeed coeval, validating a large body of published work that relies on that to be the case. Furthermore, our results provide strong evidence that chemical clocks indeed carry important information about stellar birthplaces and chemical evolution, and thus we propose that including them in chemical tagging efforts may facilitate the identification of nowadays dissolved stellar groups.

\end{abstract}

\keywords{stars: abundances- binaries - solar-type - solar neighbourhood}

\section{Introduction}
\label{sec:intro}

During recent years, several studies have found in FGK low-mass stars that there are some abundance ratios like [Y/Mg], [Y/Al] and [Ba/Mg] that follow a linear relation with stellar age \citep{daSilva2012, Nissen15,Feltzing17, Titarenko19, DelgadoMena19}. A way to understand why this happens is explained in \cite{Nissen15}. It is based on the fact that different nucleosynthesis processes occur at specific stages of the stellar evolution depending on the masses of the stars. One can observe, for example, that the abundance of elements that are produced in core-collapse supernovae from massive stars, like $\alpha$-elements \citep{Nomoto13}, has a very different production rate with time compared with those produced in late AGB stages of intermediate mass stars, like neutron-capture elements \citep{Karakas14}. Because these two processes enrich the interstellar medium (ISM) at different timescales, we now see a dependence with stellar age when comparing the two families of elements. 

Following this idea of different chemical enrichment rates for elements of different nucleosynthetic channels,  \cite{Jofre20} looked for more trends in abundance ratios between different nucleosynthetic families using a large sample of solar twins analysed by \cite{Bedell18}. They found that 55 different abundance ratios have similar {\it chemical clock}-behavior. All these ratios involved neutron-capture elements (s-process produced by AGB mostly). These abundance also have shown to add crucial information in chemical evolution studies for metal-poor stars \citep[see][] {Skuladottir19}. 

Although chemical clocks have been the subject of many studies, their applications and limitations are not yet fully understood. In particular, \cite{Feltzing17} challenged the chemical clock applicability of [Y/Mg] by showing that it depends on metallicity. Such dependencies have been further studied for [Y/Mg], [Y/Al] by \cite{Casali20} and [Sr/Mg] by  \cite{Nissen20} showing a significant effect. \cite{Casamiquela21} used a large sample of open clusters with well-known ages, to show that offsets and dispersion are introduced in the age-abundance relations when reaching out of the local bubble ($d>1$ kpc).
Therefore, it is not straightforward to apply these relations outside the solar-metallicity solar-vicinity.\\ 

Another test that could be performed to evaluate the applicability of chemical clocks would be to study how consistent they are between stars that were born together, with systems that allow covering a much greater range of ages and metallicities.
Wide binary systems  ($ 100 \text {AU} <a <1 \text {pc} $) meet these conditions, since their possible formation scenarios, such as the association of pairs of stars during the dissolution of their cluster of stars birth of a single age \citep{Kouwenhoven10, Penarubia21}, dynamic deployment of triple systems \citep{Elliott16}, turbulent fragmentation \citep{Lee17} and gravitational attraction of nearby prestellar nuclei \citep{Tokovinin17}, imply that they are formed by stars which have a common origin, being an open cluster analog of only two stars. As a consequence, wide binaries are expected to have similar chemical compositions and ages. To our knowledge, the only work that studies the latter with a large sample of systems is that of \cite{Kraus09}, where they confirmed that young binary stars in Taurus-Auriaga association are coeval within $\sim 0.16 $ dex in $\log \tau$.

Originally proposed by \cite{Andrews18, Andrews19}, wide binaries are an ideal sample for studying Galactic chemical tagging \citep{Freeman02}, because the consistency in the detailed chemistry  of their components (e.g. abundances and metallicity) imply a common origin for these stars. In addition, wide binaries span a wider range of age and $[\text{Fe/H}]$ than open clusters, thus making it possible to understand the possibility of chemical tagging over a more representative fraction of the Galactic stellar populations.  Several studies have already analyzed chemical abundances in wide binary systems \citep[e.g.][]{Oh17, Ramirez19, Hawkins20} and found that they are very homogeneous (within $\sim 0.1\text{ dex}$ and $\sim 0.05\text{ dex}$ for stars with similar $T_{\text{eff}}$), but also found that there are some cases in which these systems seem to have statistically significant differences in chemical composition. If the chemical abundances follow a trend with condensation temperature, this can be typically attributed to planetary formation or planet engulfment \citep{Melendez09, Teske15, Maia19}, enhancing the abundances of refractory and/or Fe-peak elements. \cite{Hawkins20} used a sample of 25 wide binaries such that their components were of similar spectral type and found that odd-z elements, Fe-peak, $\alpha$-elements, and neutron-capture elements are very consistent, in contrast to what is found when random pairs are used. To test the applicability of chemical tagging, high precision abundances are key. 

Such high precision abundance analysis can be done by performing line-by-line differential analyses using a star with similar stellar parameters as reference, like the Sun for solar-twins. This helps to minimize the effects of systematic errors in the derivation of the abundances such as approximations in the model atmospheres and/or inaccuracies in the atomic data, and thus achieve high-precision measurements \citep[see][for more details]{Jofre19, Nissen18}. These levels of precision \citep[$\sim0.01$ dex][]{Melendez09} open the door to new ways of multidimensional studies of the chemical evolution of the Milky Way \citep[e.g. using evolutionary trees][]{jofre2017_phylo,jackson20}.

The aim of this work is to evaluate how consistent the various chemical clocks are among the components of wide binary systems, taking advantage of their expected contemporary nature.
Section~\ref{sec:data} explains the data used, and the process with which we derived the stellar parameters and chemical abundances of our sample is described in Section~\ref{sec:methods}. The results regarding chemical homogeneity are explained in Section~\ref{sec:chemhom}, and in Section~\ref{subsection:chemclocks} we present a detailed analysis of the consistency found in chemical clocks. Finally, our conclusions are in Section~\ref{sec:conclusion}.

\section{DATA} \label{sec:data}

For this study we used a sample of 36 wide binary systems in total. Among them, 5 were observed and analyzed by us, while the rest was compiled from different sources in the literature. This allowed us to have better statistics in our results.

\subsection{Our sample of wide binaries}

Our 5 wide binaries have solar-type components and have been previously classified as bonafide binaries using common proper motions and photometry \citep{Chaname04,Gould04}.
We checked their parallaxes ($\varpi$) from and radial velocities from \textit{Gaia} EDR3 \citep{DR3}, to confirm that the pairs belong to a wide binary system. We also computed the projected separation (s) of each system, considering the inverse of the mean parallax as distance and calculating their angular separation $\theta$, as done in \cite{Andrews18} (see Table~\ref{table:obsdata}). Given the consistency in these parameters we can say that they are bonafide wide binaries.

\begin{deluxetable*}{llrrrrrr}
\tablenum{1}
\tablecaption{Right ascension ($\alpha$), declination ($\delta$), observational data, and projected separation of our sample of five wide binaries}\label{table:obsdata}
\tablewidth{0pt}
\tablehead{
\colhead{Star} & \colhead{$\alpha$} & \colhead{$\delta$} & \colhead{$\varpi$} & \colhead{Date} & \colhead{Exp time} & \colhead{S/N} & \colhead{$s$} \\
\colhead{} & \colhead{(deg)}  & \colhead{(deg)}  & \colhead{(mas)} & \colhead{(s)} & \colhead{} & \colhead{} &\colhead{(AU)}}
\startdata
 HIP32871 & 102.75 & -56.24 & 13.32 & 2011 Sep 09 & 1711.9 & 344 & 1254.51 \\
 HIP32865 & 102.74 & -56.24 & 13.31 & 2011 Sep 09 & 2473.9 & 342 &  \\
  \hline
 HIP34426 & 107.05 & 15.52 & 20.97 & 2012 Jan 23 & 2615.4 & 506 & 8218.91 \\
 HIP34407 & 107.00 & 15.53 & 21.02 & 2012 Jan 23 & 1800.0 & 409 &  \\
 \hline
 HIP52792 & 161.90 & -15.63 & 12.08 & 2011 Nov 04 & 365.9 & 332 & 2594.17\\
 HIP52793 & 161.90 & -15.62 & 11.56 & 2011 Nov 04 & 1911.8 & 355 &  \\
 \hline
 HIP58240 & 179.18 & -32.27 & 28.23 & 2012 Feb 22 & 1787.6 & 299 & 664.31 \\
 HIP58241 & 179.18 & -32.27 & 28.21 & 2012 Feb 22 & 1440.0 & 297 &  \\
 \hline
 HIP15304 & 49.36 & 7.66 & 21.27 & 2011 Sep 08 & 661.5 & 310 & 7374.10 \\
 HIP15310 & 49.39 & 7.69 & 21.23 & 2011 Sep 08 & 792.0 & 288 &  \\
 \hline
VESTA & 318.48 & -22.89 & - & 2011 Nov 06 & 360.0 & 255 &  \\
\enddata
\tablecomments{
\hspace{1mm} $s = \theta \times 1/\varpi $, 
\hspace{1mm} $\theta = \sqrt{(\alpha_{A}-\alpha_{B})^{2}\cos \delta_{A}\cos \delta_{B}+(\delta_{A}-\delta_{B})^{2}} $ \\
 $[\text{M/H}] = [\text{Fe/H}] + \log(0.638f_{\alpha} + 0.362)$, 
\hspace{1mm} $\log f_{\alpha}= $ [$\alpha$/H]}
\end{deluxetable*}

These five systems were observed between September 2011 and January 2012\footnote{CN2012A-063 and CN2012A-064 CNTAC programs} with the high-resolution Magellan Inamori Kyocera Echelle \citep[MIKE;][]{Bernstein03} spectrograph on the $6.5$ m Magellan Clay telescope located at Las Campanas Observatory. It has a resolution of $R\sim65,000$ for the red, and $R\sim85,000$ for the blue orders. The observations were carried using a slit of $0.35''\times5.00$ and a binning of $1\times1$. In addition, we considered a solar spectrum taken from VESTA, also observed in that period. 

The spectra cover a range of $3380\text{\AA}$ to $9370\text{\AA}$ and have signal-to-noise (S/N) above 250, which was measured between 777.65 nm and 777.95 nm, because it is a region free from absorption lines. Data  reduction  was  done  using  the MIKE Carnegie-Python  pipeline \citep[][]{Kelson03, Kelson00}

\begin{figure}
\center
\includegraphics[width=0.47\textwidth]{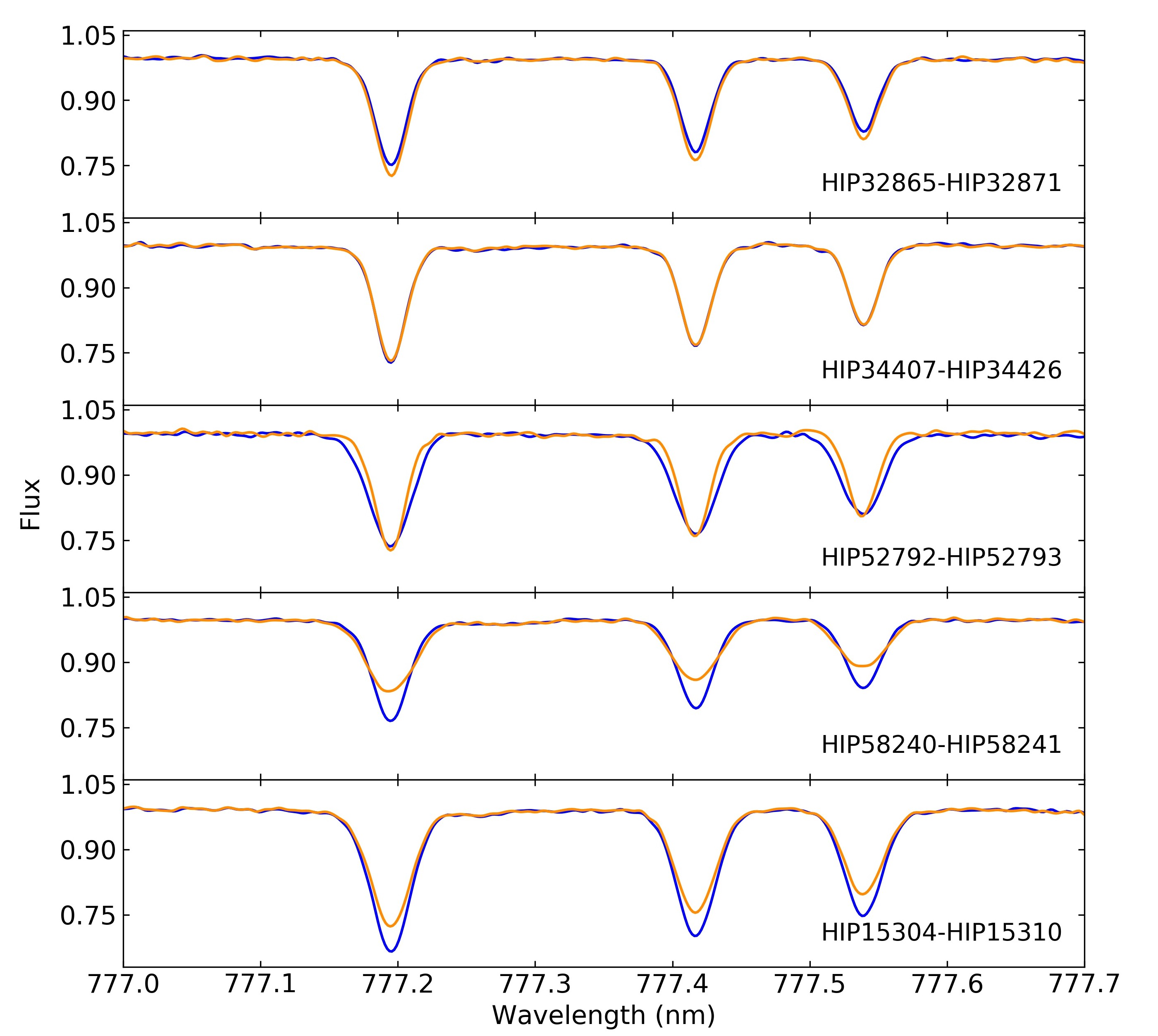}
\caption{One-dimensional spectra of our sample of wide binaries, centered in the oxygen triplet lines at $\sim$777 nm. The primary component is shown in blue and the secondary in orange.}\label{fig: spectral comparison}
\end{figure}

Figure~\ref{fig: spectral comparison} shows the oxygen triplet of our five wide binaries, where the primary and secondary component are in blue and orange, respectively. Here we see that some components have differences in shape and/or strength of their lines, which besides being caused by abundance differences can be due to line broadening (e.g. in HIP 58240/HIP 58241).

\subsection{Wide binaries in Literature}\label{sec:wbliterature}

To improve the statistics of our analysis of chemical clocks, we include 32 wide binary systems for which high-resolution spectral analysis have been performed in the literature, and abundances or equivalent widths (EWs) are available. More information about these systems, like stellar parameters, astrometry and abundances can be found in the tables included as online material. 

For systems HAT-P \citep{Saffe17}, 
HD 20782/HD 20781 \citep{Mack14}, 
HD 80606/HD 80607 \citep{Liu18},
WASP-94 \citep{Teske16}, 
XO-2 \citep{Ramirez15}, 
HD 134439/HD 134440 \citep{Reggiani18}, and 
$\zeta$ Ret \citep{Saffe16}, we used the available EWs published by the authors to derive their abundances using the same methodology than for our own observations (explained in Section \ref{sec:methods}).

In almost all of these systems, except for the last two, it has been possible to probe that at least one of its components hosts planets.
Although the existence of planets in HD 134439/HD 134440 and $\zeta$ Ret has not been confirmed, they also present a slight difference in their chemical composition, of about 0.06 dex \citep{Reggiani18} and 0.04 dex \citep{ Saffe16} in [Fe/H], respectively. This has been interpreted as possible signs of engulfment in HD 134439/HD 134440, and the existence of a debris disk around $\zeta^{2}$ Ret \citep{Eiroa10}. 

This last system is very peculiar. \cite{Nissen20} found it is depleted in yttrium compared to other stars of similar ages. In addition to this, their isochrone and activity ages are significantly lower than the chemical ages derived with [Y/Mg] ($9.1 \pm0.5$ Gyr for $\zeta^{1}$ Ret and $9.4 \pm 0.5$ Gyr for $\zeta^{2}$ Ret), which implies that its atmosphere behaves like that of a young star, despite being chemically and kinetically old. A plausible explanation for this if that they are blue straggler stars formed by short period binaries \citep{RochaPinto02}. 

The abundances of the remaining 25 wide binaries were taken from \citet[][hereafter H20]{Hawkins20}, where they derived parameters and abundances with spectral synthesis using the BACCHUS code \citep{Masseron16}. We noted that their system called WB01 is in fact HD 80606/HD 80607\citep{Liu18}. In order to avoid counting it twice, we only kept its data from H20.

There are no studies that confirm or refute the existence of planets in our systems or those of H20 (except for WB01 and HIP 34426/HIP 34407 \citep{Ramirez19}). Despite this, in H20 they mention that the condensation temperature trend of the 5 systems with $\Delta [\text{Fe/H}]> 0.1$dex was studied and that, in some cases, there is indeed a clear correlation indicating accretion of rocky material, but they do not go further into the subject.

The advantage of including some binary systems whose components present variations in their chemical patterns, is that it will allow us to test the consistency of chemical clocks even when there have been additional factors that could alter the chemical composition of the stars, such as primordial differences (i.e. planet formation and engulfment) or stellar evolution.

\begin{figure}[t]
\centering
\includegraphics[width = 0.49\textwidth]{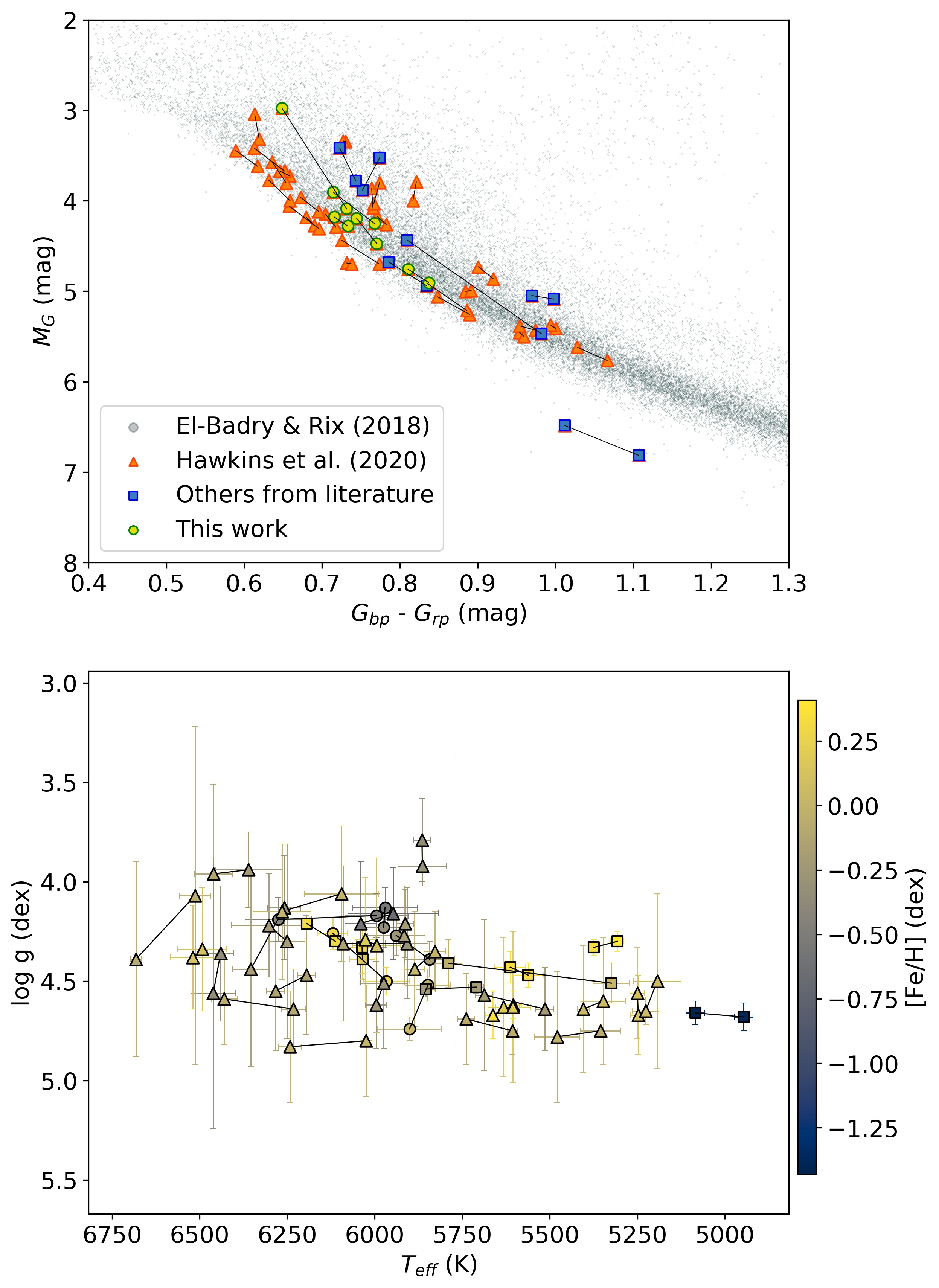}
\caption{Top: \emph{Gaia} DR3 color-magnitude diagram of the stars used in this work. Background stars (grey) were taken from \cite{EB18} catalog of wide binaries.
Bottom: Kiel diagram ($T_{\text{eff}}$ vs $\log g$) colored by metallicity. The dashed lines show the position of the Sun. HD 134439/HD 13440 is separated from the rest of the sample and lies below the main sequence, given that it is very metal-poor and suffers much less blanketing in comparison with other more metal-rich stars of the same mass.\label{fig:cmd} 
}
\end{figure}

The top panel of Figure~\ref{fig:cmd} shows our entire sample of stars in \emph{Gaia} EDR3 color-magnitude diagram (CMD) divided into 3 groups: our five wide binaries (green), H20 systems (orange), and the rest of the systems from literature (blue). For reference we show in grey circles the catalog of wide binaries from \cite{EB18}. The bottom panel shows a kiel diagram ($T_{\text{eff}}$ vs $\log g$) colored by metallicity, where the dashed lines point the location of the Sun. Each binary is connected with a black line, which allows us to see that the components have very similar colors, absolute magnitudes and stellar parameters.

Most of the pairs are well placed in the expected main sequence locus, except for HD 134439/HD 134440, which is located below. The components of this binary are very metal-poor($-1.43$ dex and $-1.39$ dex, respectively), so they are much less affected by blanketing, and therefore, bluer than a star with the same mass but higher metallicity (i.e. solar). In \cite{Reggiani18} they found that its chemical composition and kinematics are not consistent with low-$\alpha$ nor high-$\alpha$ halo stars, but instead suggest that it might have formed in a dwarf galaxy similar to Fornax. This possible intergalactic origin makes it an interesting object to test chemical clocks.

\section{CHEMICAL ABUNDANCES OF WIDE BINARIES} \label{sec:methods}

\subsection{Our sample}
The analysis of the MIKE spectra was done using the spectroscopic framework {\it iSpec} \citep{BlancoCuaresma14, BlancoCuaresma19}, which is a wrapper capable to run several state-of-the-art radiative transfer codes and model atmospheres. iSpec has also implemented several functions for basic spectra manipulation and computation of radial velocities. 

\subsubsection{Data preparation}

For each order its continuum was fitted by applying a median-maximum filter to the fluxes in order to reduce the effects of noise and ignore the deeper fluxes of absorption lines. Then, it was modeled with a third-degree spline function every $5\text{nm}$ ignoring prefixed strong lines like tellurics and strong absorption lines identified and masked using a telluric line list from iSpec. After normalization, the orders were stacked using {\ttfamily scombine} task from IRAF using the median value of the fluxes.
The one-dimensional spectra were corrected by their barycentric velocities with iSpec before determining the radial velocities from cross-correlation with a high S/N solar spectrum from NARVAL ($370 - 1048$ nm) provided by this framework.

\subsubsection{Stellar parameters}\label{subsubsec:stellab}

Both abundances and stellar parameters were determined from the observed EW, which where measured with iSpec fitting a Gaussian profile to each line. For the abundance determination, we selected the lines whose strength falls mainly in the linear part of the curve of growth \citep{Gray08}. In all our procedure we use the MARCS 1D models \citep{Gustafsson08} and LTE\footnote{Local Thermodynamic Equilibrium}  MOOG\footnote{\url{http://www.as.utexas.edu/~chris/moog.html}} radiative transfer code \citep[][2017 version]{Sneden73, Sneden12}

The stellar atmospheric parameters were determined with iSpec using the classic spectroscopic method: excitation-ionization balance. This  method requires no correlation between excitation potential and iron abundance derived from several absorption lines (excitation balance). In addition, it requires having same abundances derived from neutral and ionized iron lines (ionization balance). That way it is possible to constrain the effective temperature and surface gravity, respectively. Finally, the microturbulence velocity is obtained imposing a flat slope in the iron abundance as a function of the reduced equivalent width (REW) of each line. The errors on the stellar parameters were automatically calculated with iSpec, from the co-variance matrix constructed by the nonlinear least-squares fitting algorithm \citep[][]{BlancoCuaresma14}.

We used the extended selection of lines provided by iSpec that covers a range of $480\,\text{nm}-615\,\text{nm}$ with 154 neutral and 11 singly-ionized iron lines. 
Atomic data was taken from the fifth version of the linelist of atomic data from Gaia-ESO Survey (GES) \citep{Heiter20G} and we considered the solar abundance from \cite{Grevesse07}. 
Finally, we discarded outlier lines using a robust regression of 0.9 with iSpec, to prevent any strong dispersion in the data. Our resulting stellar parameters are listed in Table~\ref{table:stellarparam}. 

The system HIP 34426/HIP 34407 is the only one of our sample that has been spectroscopically analyzed in the literature. \cite{Ramirez19} determined their stellar parameters using a differential approach with the Qoyllur-Quipu ($q2$) code\footnote{\url{https://github.com/astroChasqui/q2}} \citep{Ramirez14} and obtained $T_{\text{eff}} = 5925 \pm 12$ K, $\log g = 4.33 \pm 0.03$ dex and [Fe/H] = -0.367 $\pm$ 0.010 dex for HIP 34407, and $T_{\text{eff}} = 6007 \pm 8$ K, $\log g = 4.32 \pm 0.02$ dex and [Fe/H] = -0.544 $\pm$ 0.006 dex for HIP 34426. These values are in good agreement with our results within the uncertainties. Also, they estimated that their ages are $6.5\pm0.8$ Gyr and $6.6\pm0.8$ Gyr, respectively, assuming that both stars have identical internal metallicities of about -0.28 dex.

\subsection{Chemical abundances}\label{chemab}

Chemical abundances were determined from EWs using an extended selection of lines from different chemical species provided by iSpec. This list is adapted for the model atmospheres, radiative transfer code and linelist we use (see Section \ref{subsubsec:stellab}).

\begin{deluxetable*}{lrrrrrrr}
\tablenum{2}
\tablecaption{Radial velocity, effective temperature, surface gravity, metallicity, and microturbulence velocity of our sample of five wide binaries obtained with iSpec}\label{table:stellarparam}
\tablewidth{0pt}
\tablehead{
\colhead{Star} & \colhead{RV} & \colhead{$T_{\text{eff}}$} & \colhead{$\log{g} $} & \colhead{[M/H]} & \colhead{$v_{mic}$}\\
\colhead{} & \colhead{($\text{km s}^{-1}$)} & \colhead{(K)} & \colhead{(dex)} & \colhead{(dex)} & \colhead{(km s$^{-1}$)}}
\startdata
 HIP32871 & 0.25$\pm$0.26 & 5938 $\pm$ 82 &  4.27 $\pm$ 0.10 & -0.24 $\pm$ 0.01 & 1.21 $\pm$ 0.04 \\
 HIP32865 & 0.0$\pm$0.24  & 5843 $\pm$ 72 &  4.39 $\pm$ 0.09 & -0.23 $\pm$ 0.01 & 1.01 $\pm$ 0.04 \\
 \hline
 HIP34426 & -12.15$\pm$0.28 & 5970 $\pm$ 93 &  4.13 $\pm$ 0.10 & -0.55 $\pm$ 0.01 & 1.26 $\pm$ 0.06 \\
 HIP34407 & -12.64$\pm$0.24 & 5974 $\pm$ 73 &  4.23 $\pm$ 0.10 & -0.34 $\pm$ 0.01 & 1.26 $\pm$ 0.04 \\
 \hline
 HIP52792 & 29.63$\pm$0.42  & 6275 $\pm$ 94 &  4.19 $\pm$ 0.11 & -0.43 $\pm$ 0.02 & 2.03 $\pm$ 0.05 \\
 HIP52793 & 28.99$\pm$0.28  & 5994 $\pm$ 91 &  4.17 $\pm$ 0.11 & -0.47 $\pm$ 0.02 & 1.33 $\pm$ 0.04 \\
 \hline
 HIP58240 & 6.21$\pm$0.22 & 5848 $\pm$ 65 &  4.52 $\pm$ 0.08 & -0.01 $\pm$ 0.02 & 1.41 $\pm$ 0.04 \\
 HIP58241 & 7.0$\pm$0.32  & 5901 $\pm$ 92 &  4.74 $\pm$ 0.06 &  0.03 $\pm$ 0.02 & 2.00 $\pm$ 0.08 \\
 \hline
 HIP15304 & 31.54$\pm$0.24 & 6120 $\pm$ 42 &  4.26 $\pm$ 0.08 &  0.22 $\pm$ 0.09 & 1.62 $\pm$ 0.08 \\
 HIP15310 & 32.18$\pm$0.24 & 5966 $\pm$ 66 &  4.50 $\pm$ 0.07 &  0.25 $\pm$ 0.08 & 1.35 $\pm$ 0.03 \\
 \hline
 VESTA & -5.26$\pm$0.2 & 5799 $\pm$ 66 & 4.49 $\pm$ 0.09 & -0.01 $\pm$ 0.07 & 1.07 $\pm$ 0.04 \\
 \enddata
\end{deluxetable*}

The linelist and the radiative transfer code include a treatment for hyperfine structure splitting (HFS) for Sc I, V I Mn I, Co I, Cu I, Ba II, Eu II, La II, Pr II, Nd II, Sm II. Also, it considers isotopic shifts for Zr I and Nd II.
That way we measured abundances of 23 different elements from the following nucleosynthetic families: carbon, $\alpha$ (O, Mg, Si, Ca, Ti), neutron-capture (Sr, Y, Zr, Ba, La, Ce), odd-Z (Na, Al, Sc, V, Cu), and iron-peak elements(Cr, Mn, Fe, Co, Ni, Zn). We use the same nucleosynthetic classification as in \cite{Jofre20}. Carbon abundances were determined using C I lines. Oxygen abundances were determined using the oxygen triplet at 777 nm, and non-LTE (NLTE) corrections where extracted from \cite{NLTE_MPIA}. These corrections typically ranged between $-0.64$ and $-0.9$ dex, with variations smaller than 0.1 dex within the components of each system. 

Once individual line absolute abundances are measured, strictly line-by-line differential abundances can be computed. For the element X, the differential abundance $\Delta X$ is defined as the mean value of the difference in abundance of each line with respect to the same lines in the reference star:

\begin{equation}
    \Delta X = \frac{1}{N} \sum_{i=1}^{N} (A_{X,i} - A_{X, \text{ref}})
\end{equation}

Where $A_{X,i}$ is the absolute abundance of the $i$th-line of the given star, $A_{X, \text{ref}}$ is the absolute abundance of the same line measured in the reference star, and N is the number of lines. Here, the reference star is the Sun, whose abundances are obtained from the spectrum of VESTA. 

Uncertainties in element abundances were obtained using the standard error ($\text{SE} = \sigma/\sqrt{N}$) when more than three lines were available. This error estimation method is frequently used \citep[e.g.][]{Melendez12, Ramirez15, Adibekyan16}, but it is important to recall that in the case of elements with few lines it can underestimate the real dispersion, and it is recommended to use the median with the interquartile $(q_{75}-q_{25})$ range as its uncertainty, because it is a more robust estimator of the abundances. \citep[see][]{Jofre18}. 

For elements with only one spectral line available we used a conservative error of 0.03, which is the maximum error for the elements with more than 3 lines. Also, we take into account the effects of the uncertainties in stellar parameters by computing the sensitivity $\sigma_{\text{sens}}$ for each element. Following \cite{Casamiquela20} this $\sigma_{\text{sens}}$ is the dispersion in abundances obtained when varying the stellar parameters by their uncertainties. The final uncertainty is obtained by adding in quadrature the element sensitivity and the standard error. 
\begin{equation}
    \sigma_{tot} = \sqrt{\text{SE}^{2} + \sigma_{\text{sens}}^{2}}
\end{equation}

\subsection{Abundances of wide binaries from the literature}

The abundances of systems collected from the literature (except H20) were determined from their EW list using the stellar parameters reported by the authors (which were derived differentially from excitation-ionization balance), and following a similar method as ours (Section~\ref{chemab}). In addition, the authors of the analyses of the seven systems taken from the literature reported the EWs of the reference star that they used (usually VESTA, except for HD 134439/HD 134440 where they used HD 103095). We used this information to perform the differential analysis in the same way as for our systems. Something important to recall is that each of these systems were observed within a different spectral range, therefore their abundances were determined with different lists of lines.
Also, we placed the abundances of H20, sample to \cite{Grevesse07} scale for consistency. 
Our final sample of wide binaries consists of 36 systems (72 stars) with stellar parameters that span a range of $\sim 4946 - 6682$ K in $T_{\text{eff}}$, $\sim 3.79 - 4.83$ dex in $\log g$ and $\sim -1.43 - 0.41$ dex in [Fe/H].

\section{CHEMICAL HOMOGENEITY OF WIDE BINARIES}\label{sec:chemhom}

The chemical homogeneity of stars in wide binary systems has been previously studied \citep[e.g.][]{Andrews18, Hawkins20}, and these components have generally been found to exhibit very similar chemical patterns, supporting evidence that they have a common origin.

\subsection{Chemical homogeneity in our systems}\label{r1}

\begin{figure*}
\centering
\includegraphics[width = \textwidth]{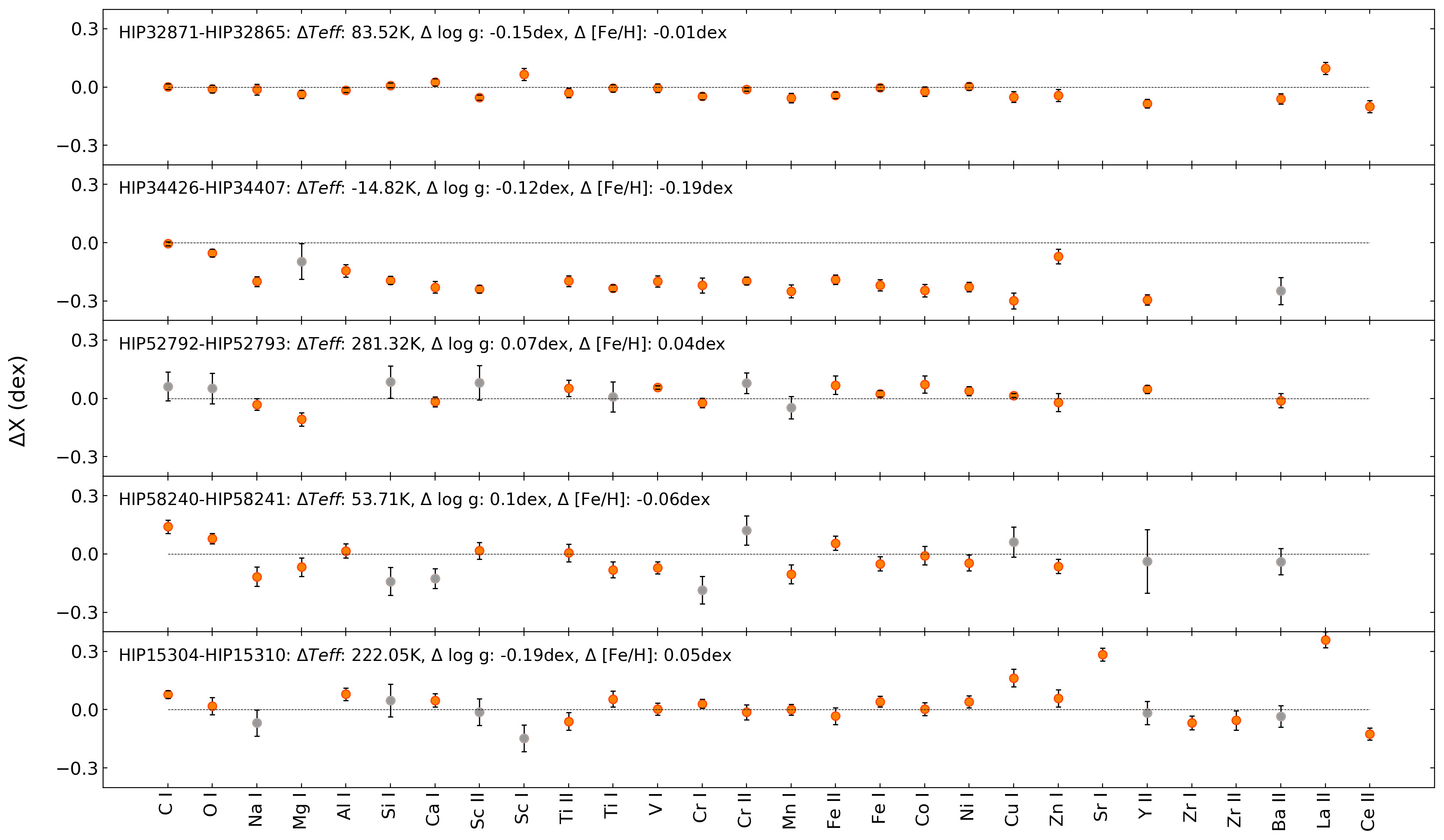}
\caption{Differential chemical abundances of wide binary pairs with solar-twin components. Each panel includes their difference in $T_{\text{eff}}, \log g$ and [Fe/H]. Grey points are abundances with uncertainties larger than 0.05 dex, whereas orange points depict uncertainties smaller than 0.05 dex. HIP 32871/HIP 32865 is a remarkable case of homogeneity, while HIP 34426/HIP 34407 has large abundance differences of 0.2 dex in average.
\label{fig:diffab}}
\end{figure*}

The abundance differences $\Delta X$ between the components of each of our systems are shown in Figure~\ref{fig:diffab}. In order to help to visualize the precision of the measurements, differential abundances with uncertainties greater than 0.05 dex are colored in grey, whereas orange points depict differences with smaller uncertainties. Each panel also includes the difference in $T_{\text{eff}}$, $\log g$ and [Fe/H] between the binary components. 

In some cases we were not able to measure all the elements listed in Figure~\ref{fig:diffab} because the same spectral line was not correctly measured in one or both components of that system. Specifically, this is the case of Zr, Sr I, Sc I, La II and Ce II, most of them n-capture elements which are known to be hard to measure since their lines are weak and sometimes blended \citep{Jofre19}. In general, $\Delta X$ is very close to zero (in average $\sim 0.1$ dex), bearing out that wide binaries systems are indeed chemically homogeneous, in agreement with the literature.

HIP 32871/HIP 32865 is a remarkable case of homogeneity whereas HIP 34426/HIP 34407 has average abundance differences of 0.2 dex in most of the elements, being the most inhomogeneous system in our sample. This system was also studied in \cite{Ramirez19}, where they proposed and analysed several scenarios that could be causing this difference in their chemical composition: (i) perhaps they are not binary from birth, (ii) engulfment of planetary material rich in refractories, and (iii) inhomogeneities in the gas cloud where they were formed. Despite the efforts, it has not yet been possible to confirm which of these scenarios led to this peculiar system.

Regarding how consistent wide binaries are in the different nucleosynthetic families, we find that the mean $\Delta X$ are $\sim$0.02 dex for neutron-capture (Sr, Y, Zr, Ba, La, Ce), $\sim$0.003 dex for $\alpha$ (O, Mg, Si, Ca, Ti), $\sim$0.03 dex for odd-Z (Na, Al, Sc, V, Cu), and $\sim$0.03 dex for Fe-peak elements (Cr, Mn, Fe, Co, Ni, Zn). This is in good agreement with \cite{Hawkins20}.

\subsection{Chemical homogeneity of the whole sample}\label{subsection:homogeneity}

\begin{figure*}
\centering
\includegraphics[width = 0.75\textwidth]{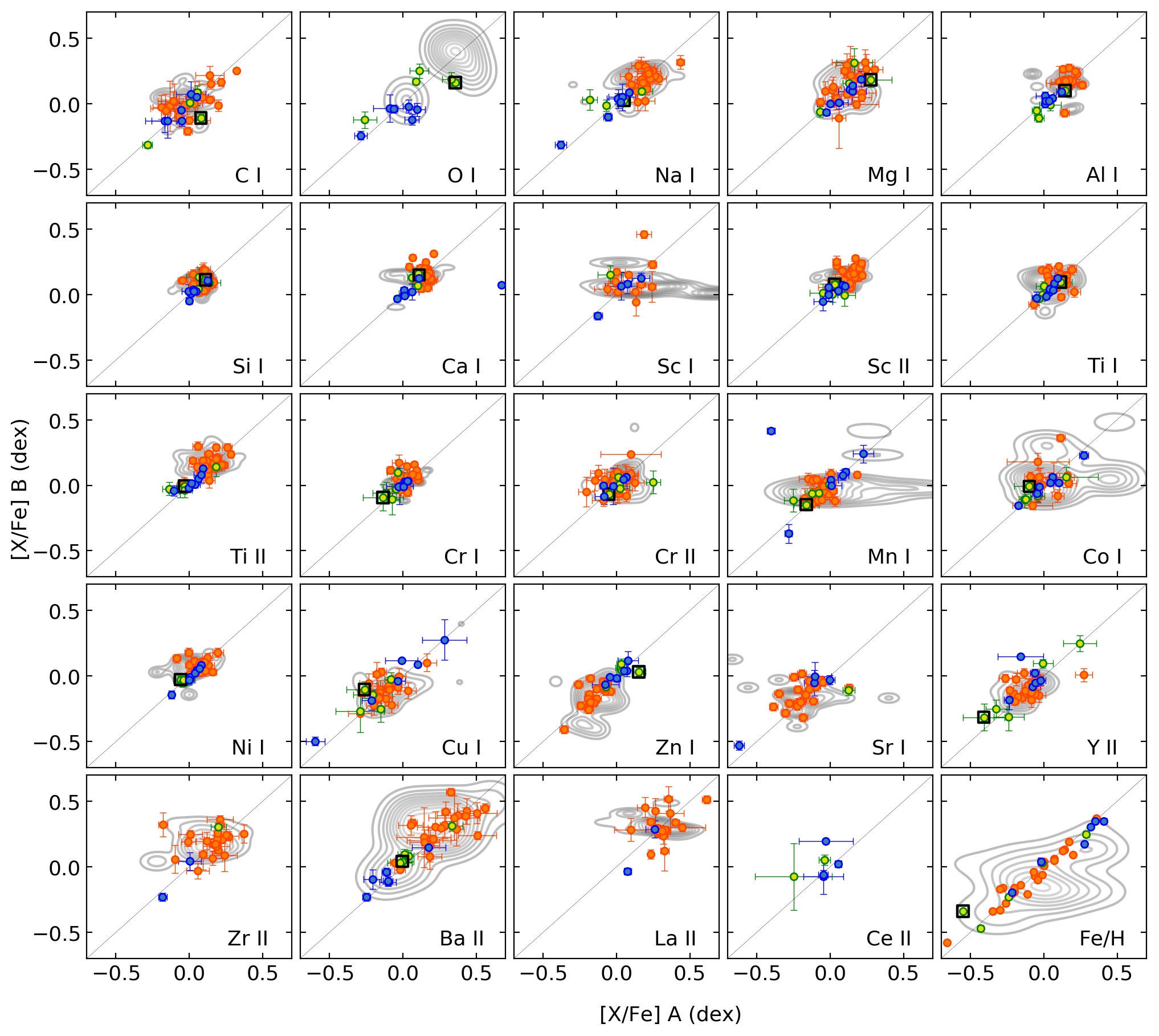}
\caption{Comparison of the individual abundances [X/Fe] of the two components of wide binaries for all the elements measured in this work. [Fe/H] is in a separated panel at the bottom right of the figure. The systems are coloured as: binaries in our sample (green), H20 (orange) and other binaries from literature (blue). The distribution of random pairs is shown as grey contour lines. If there are less than 3 random pairs with [X/Fe], their distribution is not shown. HIP 34426/HIP 34407, the most inhomogeneous system of our sample, is marked with a black square. 
\label{fig:XFeAB}}
\end{figure*}

Figure~\ref{fig:XFeAB} shows a comparison of the [X/Fe] for the 24 chemical species, for the full dataset of wide binaries. The full sample analyzed in this paper is depicted as follows: our wide binaries (green), H20 (orange) and the rest of the systems from the literature (blue). Each panel shows a different element, which is indicated at its bottom right.  [Fe/H] is included in the bottom right panel. The abundances are compared considering one component in the X axis and one component in Y axis, expecting to follow a 1-1 line if the abundances were to match exactly. 

In order to quantitatively study the degree of homogeneity of the abundances in wide binaries we compare them with the chemical behaviour of random pairs, which are not expected to follow the 1-1 line as wide binaries, given that their components were not formed together. We constructed random pairs selecting stars according to their position in the CMD (Figure~\ref{fig:cmd}) as follows. We set a star as a reference and choose the two stars (which are not its true companion) closest to it in the diagram. This is done because our sample has stars that were analyzed with different lists of lines, so it is very likely that by only selecting the closest star, there will not be an absolute match between the lines of the components of the random systems. This could considerably reduce the number of these systems in each $[\text{X/Fe}]$ and affect our statistics.
This exercise is done for each of the 72 stars of the whole sample, creating a total of 144 random systems. Grey contours show the distribution of the random pairs if at least three of them have measurements of that [X/Fe]. We note H20 does not have abundances of O or Ce. That is why some of these panels, in addition of lacking orange points, lack grey contours. 

In Figure~\ref{fig:XFeAB} we can see that wide binaries are generally more tightly distributed along the 1-1 line than random pairs, as expected from previous works \citep{Andrews18, Desidera06}.  In general, abundances of almost all of the elements show solar abundance patterns and follow a similar distribution centered at zero, possibly because these stars belong to the solar neighbourhood. 
Exceptions to this are Ba II, Mn I, Cu I, La II and Co I, which span a wider range in [X/Fe], reaching values above 0.5 dex. Regarding random pairs, they also tend to be concentrated within this range, except in some cases (Mn I, Sc I, etc). 

We further see that some elements show systematic differences among the three samples in the sense that they seem to cover different abundance regimes, e.g. Mn I, Sc I, Ba II, Co I. The reason behind this might be that each of these groups were analyzed in different ways (see Section \ref{sec:wbliterature}), or NLTE effects. Different methodologies, even when using the same prescription and input data, can lead to systematic differences in abundances. In this case, the fact that H20 abundances where determined with spectral synthesis, whereas EWs were implemented in the rest of the sample can lead to inconsistencies given the different HFS treatment of each method, as discussed in Section \ref{subsec:systematics}. Also, an extensive discussion in this subject can be found in \cite{Jofre17}. 

According to the excitation-ionization balance, neutral and ionized abundances of an element should be equal, but in practice this is not always the case given NLTE, 3D effects or HFS \citep[see][]{Jofre15}. We have 3 elements with abundances for the neutral and ionized species: Sc, Cr, Ti. Previous studies have found that Sc I  and Cr I are strongly affected by NLTE \citep{Zhang08, Bergemann10, Battistini15}, while for Ti I the corrections are minor, but it is still more sensitive to NLTE \citep{Bodaghee03}. It is for this reason that we decided to  only use the Sc II, Cr II and Ti II measurements as the abundances of each element for the rest of the following analysis.

It's important to mention that our results for elements like Na, Al, Mn, Co and Ba may be slightly affected by NLTE effects, but since we are performing a differential analysis these are minor corrections, these should mostly be represented as translations along the 1-1, not as variations in the dispersion around it. Regarding 3D effects, given that their implementation provides absolute abundances with greater precision \citep[e.g.][]{Nissen18}, it is absolutely necessary to include them when calibrating chemical clocks, but in the case of this work where we only compares the clocks of stars with similar stellar parameters it is not expected that it plays a big role, since the effect of the 3D models on the abundances should be very similar (if not the same) for both.\\

These results and those found in previous studies regarding the similar chemical composition of wide binaries (within $~0.1$ dex), serve as sufficient evidence that these systems are excellent laboratories for chemical tagging, studying effects such as planet presence, and also to test the consistency of chemical clocks, since they support the premise that these stars were born together.

\subsection{Condensation temperature trend}

The formation of planets can leave a subtle imprint in their host's stellar atmosphere ($\sim 0.01$ dex) \citep[]{Chambers10}.  While in general planets are formed from the same material as their host star, rocky planets in particular would preferentially remove refractory content (higher $T_{c}$), so they may leave the stars slightly poor in these elements \citep{Melendez09, Ramirez09, Nissen15}. Another scenarios in which the stellar composition can be altered are planet engulfment or accretion of material with planet-like composition \citep[e.g.][]{Saffe17, Maia19}, thus given the nature of these cases and following the same logic as before, they should also leave evidence in the abundance of refractory elements of the star 

\begin{figure*}
\centering
\includegraphics[width = 0.93\textwidth]{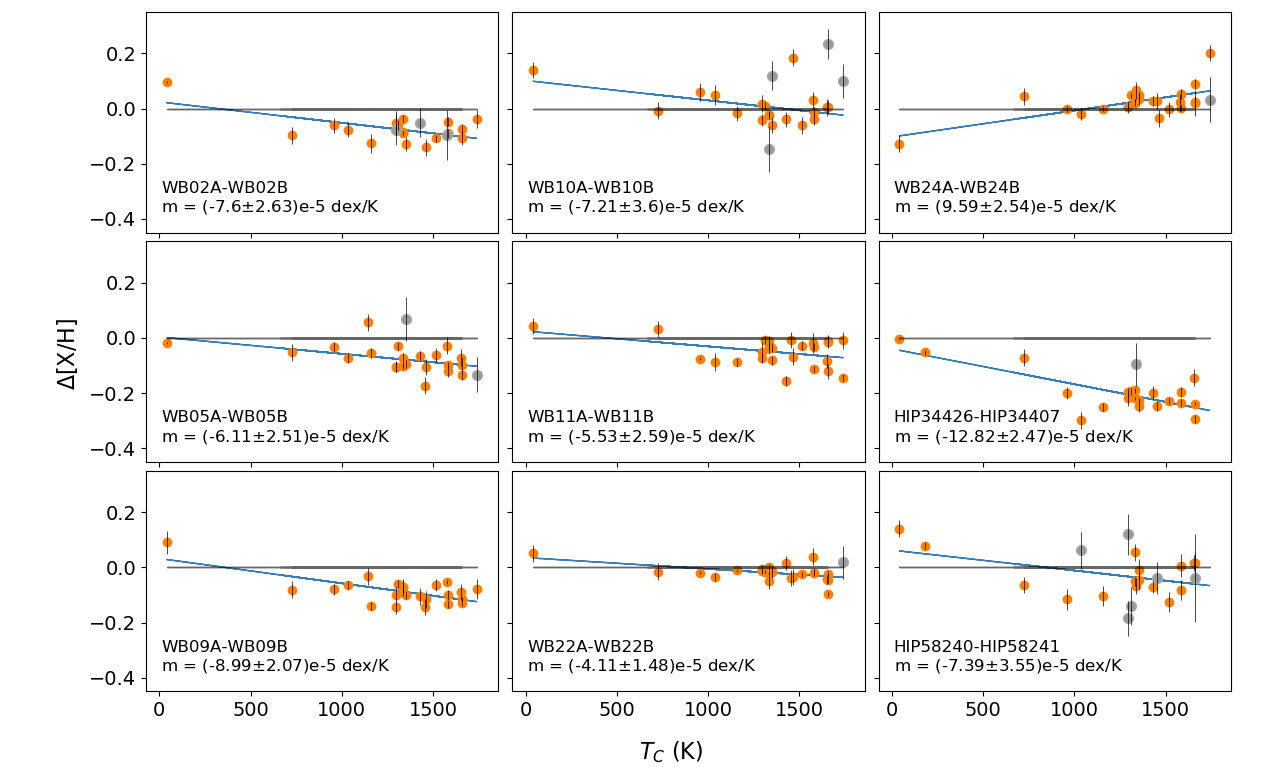}
\caption{Differential abundance as a function of elemental condensation temperature from the H20 systems and our sample of wide binaries that present a slope greater than $3\times 10^{-5}$ dex K$^{-1}$. The blue line shows the linear regression that better fits the data, which parameters are included in the legend of each system. Differential abundances are colored in the same way as in Figure~\ref{fig:diffab}. The list of elements sorted by increasing $T_{c}$ is the following: C, O, S, Zn, Na, Cu, Mn, Cr, Si, Fe, Mg, Co, Ni, Ba, Sr, Ce, Ca, La, Ti, Al, Sc, Y, and Zr. 
\label{fig:TcH20}}
\end{figure*}

Figure~\ref{fig:TcH20} shows $\Delta[\text{X/H}]$ as a function of condensation temperature, $T_{c}$ \citep{Lodders03}, for H20 and the wide binaries in our sample that present a trend with an absolute value of the slope greater that $3\times 10^{-5}$ dex K$^{-1}$ with a significance greater than 2$\sigma$. This cut is based on the slopes that have been reported in the literature, with which a trend has been confirmed or denied \citep{Mack14, Saffe16, Teske16, Liu18, Reggiani18}. The linear regression was calculated using the function {\ttfamily linregress} of SciPy library\footnote{\url{https://docs.scipy.org/doc/scipy/reference/generated/scipy.stats.linregress.html}}. Since this function does not include the individual error of the elements to derive the uncertainty of the fit, nor does it allow to make a weighted fit as suggested in \cite{Adibekyan16}, we performed the linear regression for the abundances with errors smaller than 0.05 dex (orange points in Figure~\ref{fig:TcH20}).

The linear regression that better fits the data is plotted as a blue line, and its parameters are included in each panel. We leave the other sample of systems taken from literature (blue systems in Figure~\ref{fig:cmd}) out of this plot, because the presence or absence of trend has already been studied in their respective works, and we found their slopes to be consistent with previous results.

In H20 they briefly discussed about this trend in the systems with $\Delta [\text{Fe/H}] \geq 0.1$dex, which we identify as WB02, WB05, WB09, WB16 and WB21. Here, we find that three of them, WB02, WB05, and WB21, present a considerable trend, but also WB10, WB11, WB22 and WB24. Regarding our sample, we find that HIP 34426/HIP 34407 and HIP 58240/HIP 58241 also show an important trend with condensation temperature. \cite{Ramirez19}  studied this trend in HIP 34426/HIP 34407 and found a similar behaviour. The particular chemical inhomogeneity of this system causes it to have the highest slope among all the wide binaries in this study, for which one would point to the formation or accretion of rocky planets as the main responsible for their nature \citep[as discussed in][]{Ramirez19}. Something similar could be proposed for the other systems. 
Despite this, there is still no way to know with certainty which specific mechanism is causing this correlation between chemical difference and $T_{c}$ just with the information that we handle. Probably, a systematic search for planets in this system would help understanding if the slope is related to this effect.

We noted that the element with the lowest $T_{c}$, C, seems to have a large influence on the linear regression, so we performed the fit without this element and found that only WB05 and HIP 34426/HIP 34407 keep presenting a significant trend. C I lines have negligible NLTE corrections in solar-type stars \citep[i.e.][]{Caffau10}, but only a few of them are useful to determine carbon abundances, thus molecular features are implemented. However, it is important to implement 3D atmospheric models in order to have better agreement between atomic an molecular features, as shown in \cite{Asplund05}.

\section{Chemical clocks}\label{subsection:chemclocks}

In addition to common chemistry, that we have evidenced in the previous section, the expected common origins of the components of wide binaries imply that they should have similar ages as well \citep[e.g.][]{Kraus09}.  This coevality allows us to test the reliability of the abundance ratios known as chemical clocks (Section \ref{sec:intro})
 
We follow \cite{Jofre20}, who used a sample of solar twins to present a large number of combinations of chemical abundance ratios shown to have strong dependency with ages. These relations can be explained with similar arguments as [Y/Mg], which is widely discussed in the literature \citep[e.g.][]{Nissen15,DelgadoMena19, Casamiquela21}. The argument is the very different timescales for the production of an alpha-capture element such as Mg in comparison with a neutron-capture element such as Y. When combined, a tight relation of [Y/Mg] can be found for stars of the same metallicity but different ages. \cite{Jofre20} and \cite{Casamiquela21} found that the clocks showing largest age dependency involved neutron-capture elements with elements produced by other nucleosynthesis channels, as expected given their production scenarios.

\begin{figure*}
\centering
\includegraphics[width = \textwidth]{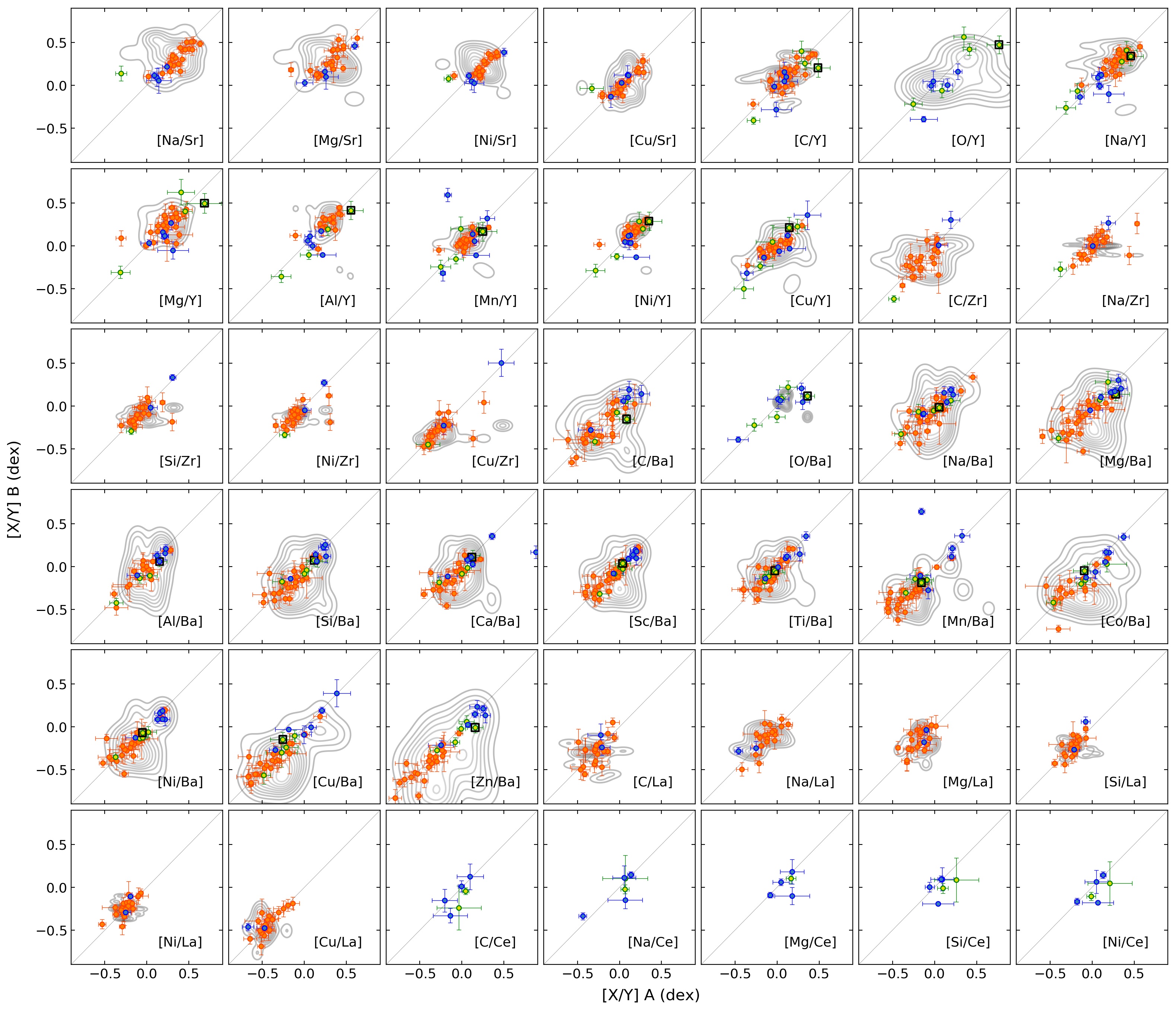}
\caption{Comparison between the components of wide binaries for 42 potential chemical clocks considered in this work. The different systems are coloured like in Figure~\ref{fig:XFeAB} and the distribution of random pairs is shown in grey curves. Same as in Figure~\ref{fig:XFeAB}, HIP 34426/HIP 34407 is marked with a black square for easy visual identification. The ages related to chemical clocks increase from left to right, so that older stars have more positive values and younger stars more negative values.
\label{fig:chemclock}}
\end{figure*}

We recovered 42 of the abundance ratios\footnote{[X/Y] = [X/H] - [Y/H]; $\delta$[X/Y] = $\sqrt{\delta X^{2} + \delta Y^{2}}$} found in \cite{Jofre20}. Figure~\ref{fig:chemclock} is similar to Figure~\ref{fig:XFeAB} but we plot chemical clocks instead of [X/Fe] abundance ratios. In each axis we plot each component of the binary, expecting the values to follow a 1-1 line.  Furthermore, we included the distribution of random pairs, like in Figure~\ref{fig:XFeAB}. Since H20 lacks of Ce and O abundances, there are 8 chemical clocks for which we are unable to have an estimate of a random pair correlation.  

In addition, we highlight the system HIP 34426/HIP 34407, which is the least homogeneous wide binary of our sample (see individual results in second panel of Figure~\ref{fig:diffab}). This system is marked with a black square, and despite being really inhomogeneous this does not seem to have an effect on some of its chemical clocks, especially those that depend on Ba (e.g. [Ti/Ba], [Al/Ba], [Ni/Ba]). Some chemical clocks span a wider range in abundance ratio than others, like  [Cu/Ba], [Zn/Ba], [Na/Sr], and [O/Y], particularly because of the H20 sample. As well as in Figure~\ref{fig:XFeAB}, we believe that this is mostly due to a systematic effect in the derived abundances, as mentioned in Section \ref{subsec:systematics}.

\subsection{Consistency of chemical ages in wide binaries}

\begin{figure}
    \centering
    \includegraphics[width = 0.38\textwidth]{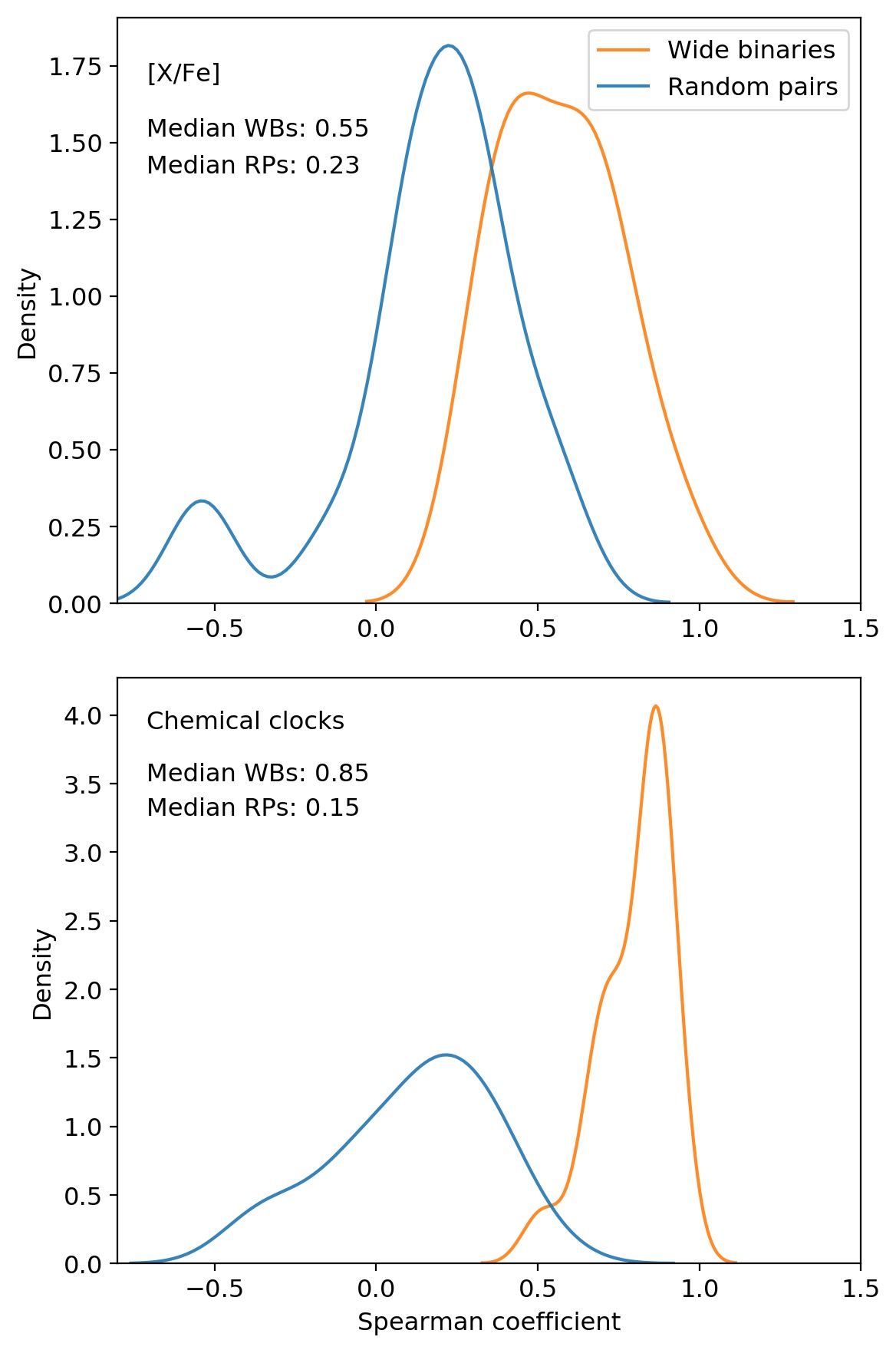}
    \caption{Distribution of the Spearman correlation coefficient (SCC) from  Figures~\ref{fig:XFeAB} (top) and \ref{fig:chemclock} (bottom), colored in orange for wide binaries and in blue for random pairs. Statistical features of each plot are listed in Table~\ref{table:sccstats}.}
    \label{fig:scoeff}
\end{figure}

In order to quantify the degree of correlation in our data, or lack of thereof, we compute the Spearman correlation coefficient (SCC), which measures of the linear correlation between two data sets and was chosen given that the chemical clocks in this work not necessarily follow a normalized distribution \footnote{Derived with SciPy: \url{https://scipy.org}}
for both wide binaries and random pairs for clocks and standard abundance ratios. In Figure~\ref{fig:scoeff} we plot the distributions of the SCC from Figures~\ref{fig:XFeAB} and \ref{fig:chemclock}, showing wide binaries in orange and random pairs in blue. The coefficients were derived only from those chemical clocks and [X/Fe] ratios with at least 10 random pairs, to ensure robust statistics. Table~\ref{table:sccstats} presents the median and median absolute deviation (MAD) of the SCC of chemical clocks and individual abundances in wide binaries and random pairs, for better understanding of Figure~\ref{fig:scoeff}.

The top panel of Figure~\ref{fig:scoeff} shows that standard [X/Fe] abundances are more consistent in wide binaries than random pairs, but the differences between distributions are not so stark. On the other hand, in the bottom panel we can see that the SCC distributions of chemical clocks are very different from each other, since random pairs show a peak at $\sim 0.1$ while in wide binaries this is at $\sim 0.8$ (see Table~\ref{table:sccstats}). Moreover, the areas covered by each distribution almost do not overlap, which means that there is a very low chance that stars that do not necessarily have a common origin can have the same chemical clocks, thus age. 

If chemical clocks were just reflecting abundance ratios, their 1-1 correlation should be comparable to the correlation found for other individual abundances between coeval stars that were born together. However, our results imply that chemical clocks are even more consistent, therefore, they do not only trace chemical homogeneity among wide binaries, they must  be carrying age information as well.
Moreover, since our sample included some systems which composition has been subtly affected by different mechanisms (e.g. planet formation, planet engulfment, etc), Figure~\ref{fig:scoeff} helps us to confirm the idea that chemical clocks are sensitive to the age of the stars regardless of the possible differences in their compositions, and still be consistent between the components of such coeval systems. The reason behind this is that the effect in refractory elements is very small, and affects them almost equally, with a scatter that is generally less than 0.03 dex \citep[e.g.][]{Maia19}. So, removing this signature from the abundance of the star and deriving the chemical clocks would lead to variations of less than 0.03 dex.

\begin{deluxetable}{llrr}
\tablenum{3}
\tablecaption{Median and mean absolute deviation (MAD) of the SCC calculated for chemical clocks and [X/Fe], for wide binaries (WBs) and random pairs in Figure~\ref{fig:scoeff}. 
\label{table:sccstats}}
\tablewidth{0pt}
\tablehead{
\colhead{} &\colhead{} & \colhead{Median} & \colhead{MAD}}
\startdata
 Chemical clocks & WBs & 0.85  &  0.04\\
 { } & random pairs & 0.15 & 0.18 \\
 \hline
 [X/Fe] & WBs  & 0.55 & 0.13\\
 { }   & random pairs & 0.23 & 0.13\\\enddata
\end{deluxetable}

In a different scenario, if we simply consider true that chemical clocks are capable of identifying contemporary systems, our results would confirm that the components of wide binaries are coeval, and it would be one of the first works in literature to do an analysis with such a large and varied sample of them.\\

\subsection{The case of  HIP 344026/HIP 34407}
In previous sections we have checked that stars in wide binaries have homogeneous chemical compositions within the uncertainties. This confirms expectations and previous results \citep{Andrews18, Hawkins20} since all known plausible formation scenarios indicate that they share common origins, hence they must share their chemical composition, especially if they are on somewhat similar evolutionary states. 

All that said, however, the case of HIP 34426/HIP 34407 is puzzling and is worth discussing. From Figure~\ref{fig:diffab} we see that abundances are very different (about 0.2 dex), with C and O the only exceptions. These stars were already analysed by \cite{Ramirez19}, who suggest the differences might be coming from stars not being truly siblings but chance alignments, the birth cloud not being homogeneous, or engulfment of planetary material. After performing an analysis of the astrometry and the radial velocities of the system, they found that HIP 34426 shows a variation in radial velocity of approximately 60 $\text{m s}^{-1}$, something that could indicate the presence of a planet such as a large Jupiter or the existence of a third component of the system. Furthermore, the difference in the total velocity of the system positions it as a physically unbound system, but given its age, \cite{Ramirez19} propose that it was a binary at formation.

If these stars were not siblings, but instead were (1) formed at separate sites with a difference of about 0.2 dex in metallicity, or (2) sites that suffered the possible effects of chemical evolution (e.g. some considerable time span between the formation of both components) or space-phase (e.g. stellar migration), then, unless a large coincidence, they most likely would have significantly different ages and, therefore, different clocks. However, according to \cite{Ramirez19} their ages are very consistent, which most likely rules out these possibilities and that of a chance alignment, since the probability for two independent and coeval stars to become a wide binary by dynamical capture in the field is negligible. Furthermore, we see in Figure~\ref{fig:chemclock} that within the 22 chemical clocks measured in this system, 14 of them show that the system is located almost perfectly on the one-to-one line despite the seemingly remarkable inhomogeneity (for a typical wide binary system) in individual abundances [X/Fe]. 

If we take the similarity of the chemical clocks as an indication for the stars in this binary being coeval, then we interpreted our results as strong evidence that the cloud material from which they formed was inhomogeneous to these levels, on a scale equivalent to the projected separation of the components of this pair. This may have important implications for the idea of chemical tagging for Galactic archaeology.  The case of binary systems like this stands in stark contrast with the recent detailed study of \cite{Kos20}, who determined chemical abundances of about 300 GALAH stars distributed in the Orion nebula, finding that the nebula is chemically homogeneous (typically within 0.1 dex in each element, except Li), at higher level of precision than the difference observed in the HIP 34426/HIP 34407 system.

Recently, \cite{Penarubia21} proposed that the formation of ultra-wide binaries ($s > 0.1$ pc) is via chance entrapment of unrelated stars in the tidal streams of disrupting clusters, and found that most pairs formed in tidal streams can easily be disrupted by subtle gravitational interactions with other objects. Here we propose that HIP 34426/HIP 34407 might be part of a tidal stream. This scenario could explain some of the characteristics that we observe in this system, such as its coevality and the astrometric differences that in principle classify it as a non-physically bound system \citep[see][]{Ramirez19}.

\subsection{Systematic uncertainties and applicability of chemical clocks}\label{subsec:systematics}

We are aware that different spectral methodologies can lead to significantly different abundance results \citep{Hinkel16, Jofre17,Jofre19,jonsson2018}. Our analysis depends on data taken from the literature, and therefore can be affected by systematic uncertainties. 

One of them is that, in spite that we used the same linelist for our analysis for the pairs for which we had EWs, H20 derived their chemical abundances  using a different selection of lines than ours. \cite{Jofre19} extensively discusses the effects of line selections in resulting abundances.

In particular, while being consistent among all pairs in our full sample, we can see from Figure~\ref{fig:chemclock} that the ratios [Zn/Ba], [Mg/Ba], [Mn/Ba], and [Cu/Ba] have values very different for H20 than the rest of the sample: their values show a tendency to negative values, that do not overlap with range of abundances found in \cite{Jofre20}.
We note that \cite{Jofre20} used a sample of stars spanning very young to very old ages, leaving little space for a systematic offset in these abundance ratios due to age. 

\begin{figure*}
\centering
\includegraphics[width = 0.98\textwidth]{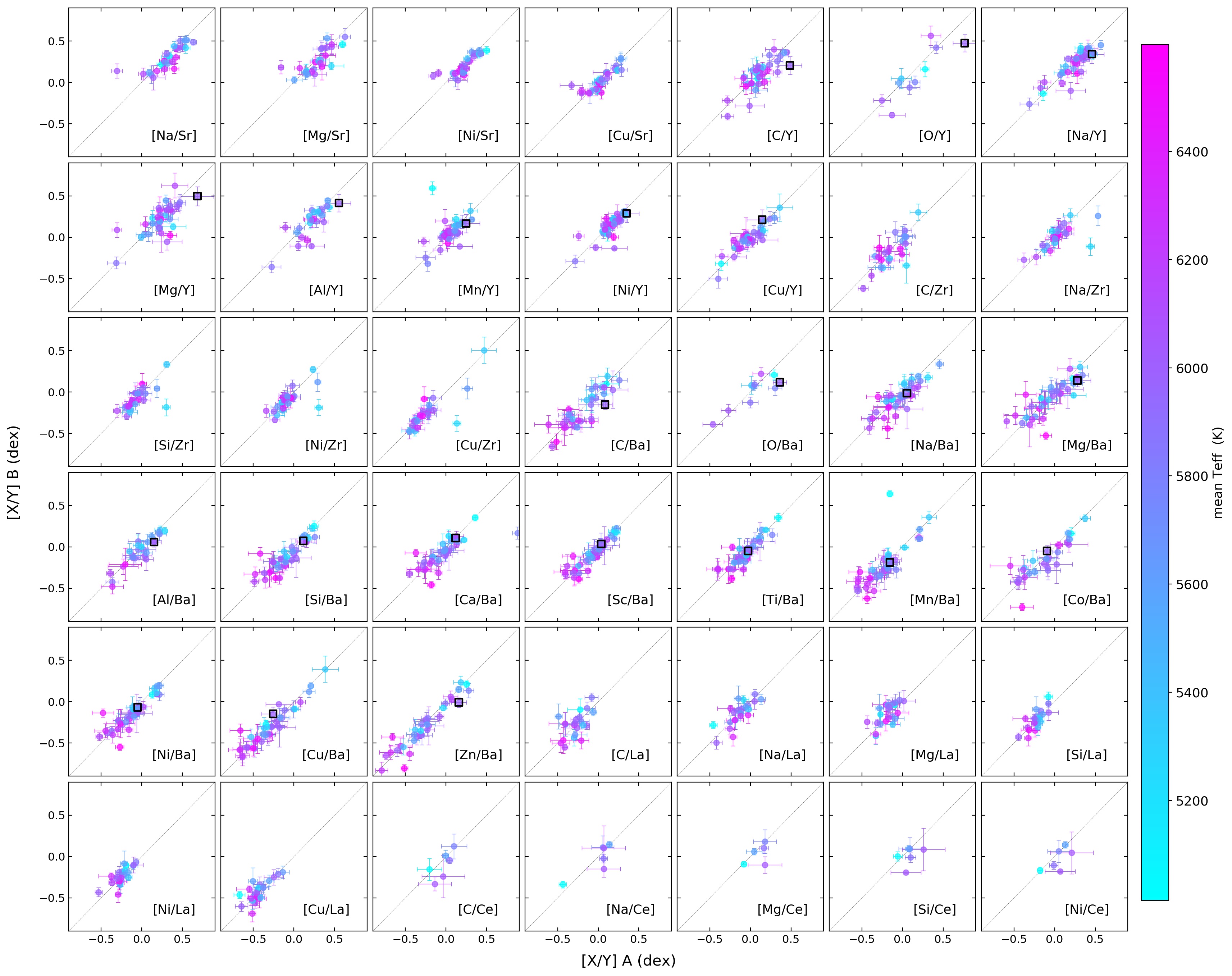}
\caption{Comparison between the components of wide binaries for potential chemical clocks considered in this work. Colors depict the mean temperature of the system. Generally, stars with higher temperatures ($\sim 6000-6200$ K) tend to produce ratios concentrated at lower absolute values than cooler stars, and hotter stars seem to be more dispersed around the 1-1 line.   \label{fig:chemclockT}}
\end{figure*}

To investigate this further, in Figure~\ref{fig:chemclockT} we plot the same abundances as in Figure~\ref{fig:chemclock} but with the pairs colored by the average $T_{\text{eff}}$ of the system. It is evident from the plot that the cases of [X/Y] with large range along the 1-1 line, also show a gradient with stellar temperature, i.e. mass. 

Generally, stars with higher temperatures ($\sim 6000- 6200$ K) tend to produce ratios concentrated at lower absolute values than cooler stars. We particularly remark the cases where Ba is involved (e.g. [Ti/Ba], [Si/Ba], [Na/Ba]), showing that there probably is some strong effect depending on the stellar mass for the retrieval of abundances of this element. Moreover hotter stars seem to be more dispersed around the 1-1 line. We recall that the hottest stars are also those analyzed by H20. 

It is worth commenting that the stellar temperature has a large impact in the line strength of neutral atomic lines (this is why we adopted the excitation balance method to determine temperature, see Section~\ref{sec:methods}). The line strength can have a particular large implication in determining abundances which have strong effects due to hyperfine structure splitting (HFS), such as Mn, Co or Ba.  
The way different methods deal with HFS can indeed be a major source of uncertainty for particular elements like Mn or Co \citep{delPeloso05, Jofre17}. For stars of different temperatures, abundances of Mn can be very different if different prescriptions for HFS are employed. We also refer to \cite{Casamiquela20} for the discussions about striking dependencies with $T_{\text{eff}}$ are seen (remarkably for Ba, Na, and Mn). 
The systematic difference in chemical clocks found here is therefore a combination of line selection, HFS treatment and stellar temperature. \\

There are some discussions in literature about the metallicity dependency on chemical clocks \citep[e.g.][]{Feltzing17, DelgadoMena19}. \cite{Casali20} found that the slopes of [Y/Mg] and [Y/Al] versus stellar age change with metallicity (by using steps of $\pm$0.2 dex in [Fe/H]), and related this to the star formation history of the Milky Way. They concluded that the position of the stars in the Galactic plane is another important parameter to take into account when using chemical clocks to determine ages (e.g. whether it belongs to the thin disk or to the thick disk), because at different locations stars can have different metallicities. Most recently, \cite{Casamiquela21} used a sample of red clump stars in 47 open clusters to study the abundance-age relation in a large spatial volume ($6{\text{ kpc}} < R_{\text{CG}}< 12{\text{ kpc}}$), covering regions outside the local bubble, unlike most studies on chemical clocks that use solar twins. By studying [Y/Mg] and [Y/Al] in more detail, they find that the clusters located outside the solar neighborhood appear to have a significantly greater dispersion in abundances compared to the local ones, without finding any dependence on metallicity because of the narrow range in [Fe/H] of their sample. They attribute this finding to galactic dynamics processes, such as radial migration of stars or clusters, which are more likely to be sampled when covering a wider range in distance and can trace different chemical evolutionary histories.

In general, the results of other studies indicate that chemical clocks have several limitations caused by the chemical and dynamic nature of our Milky Way, so one must be very careful when using them to derive ages. Even so, their conclusions do not present problems or restrictions regarding chemical clocks being consistent in coeval star systems. Little however has been discussed in terms of the effects in deriving ages using chemical clocks when such abundances are hampered by systematic uncertainties such as those discussed here. More such analyses are needed in order to fully assess the applicability of chemical clocks for wide Galactic archaeology studies.

\subsection{Most consistent clocks in wide binaries}

\begin{figure*}
\centering
\includegraphics[width =0.9\textwidth]{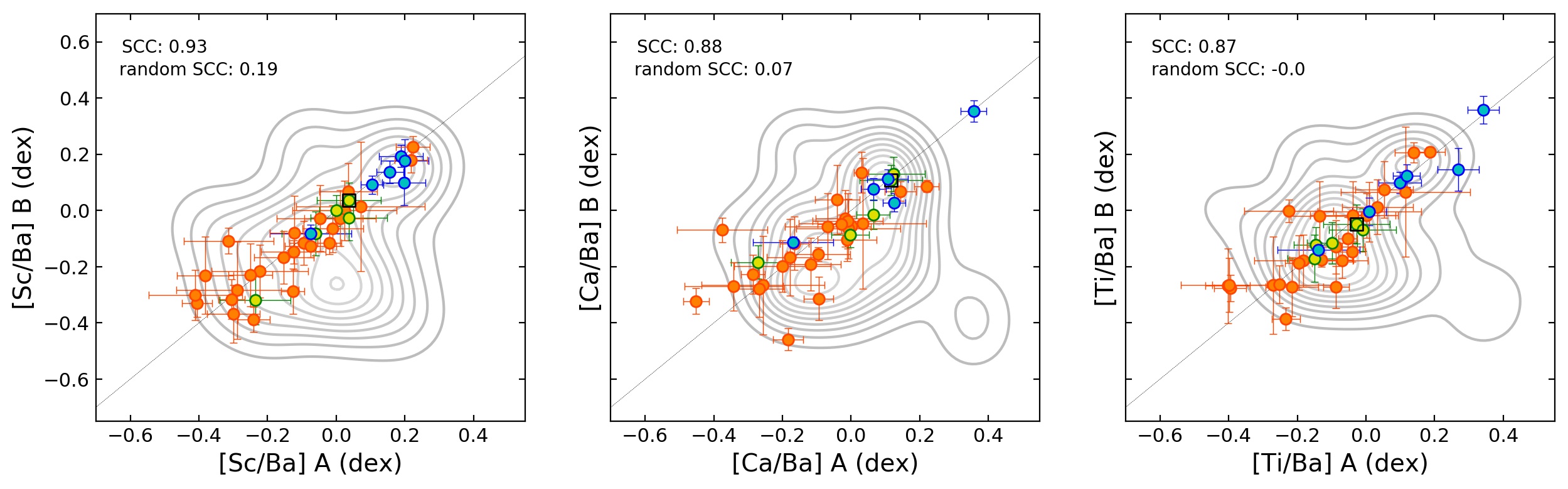}
\caption{Chemical clocks with the higher correlation in our sample of wide binaries: [Sc/Ba], [Ca/Ba] and [Ti, Ba]. The distribution of random pairs is shown in grey curves and HIP 34426/HIP 34407 is marked with a black square, as in Figure \ref{fig:chemclock}. These abundance ratios can trace the coevality of HIP 34426/HIP 34407, despite its chemical inhomogeneity. \label{fig:top3}}
\end{figure*}

Based on our results we have selected three chemical clocks that have the highest correlation coefficient for wide binaries, and that also trace the age consistency of HIP 34426/HIP 34407 system despite the chemical differences of its components. Figure~\ref{fig:top3} shows [Sc/Ba], [Ca/Ba] and [Ti/Ba] and the SCC values for wide binaries and random pairs. 

We can see that despite coming from independent spectroscopic measurements, the chemical clocks of the wide binaries fit nicely to the 1-1 line, and although we are aware that there are systematic errors in our sample due to the inclusion of systems analyzed with different methods, these do not influence when plotting the age consistency of these systems. But, if they are to be implemented in chemical tagging studies, it is necessary to follow a spectroscopic analysis as consistent as possible in order to avoid any systematic uncertainty which could make the identification of stellar groups very difficult. 

It is worth emphasizing that Figure~\ref{fig:top3} shows the potential of these abundance ratios to represent the ages of systems that arise from a common origin, even when certain chemical inhomogeneities are present, but still, it remains to be seen how these clocks relate to the individual ages of the components of wide binaries.

\section{CONCLUSIONS} \label{sec:conclusion}

In the present work we studied how consistent are certain group of abundance ratios, collectively known as chemical clocks, among the components of wide binaries. Since the latter are expected to share a common formation site and to be coeval, the comparison of their chemical clocks provides a basic check on the concept itself  of a chemical clock and its overall validity.  For this, we spectroscopically analyzed a sample of five systems, four of which for the first time, doing a differential analysis with respect to the Sun and using the equivalent widths method. We found that the components of these systems are chemically homogeneous (within 0.1 dex), in good agreement with what has been found in other wide binaries. One remarkable exception of this is HIP 34426/HIP 34407, which presents differences of $\sim 0.2$ dex in abundance of most of the elements measured. We found their chemical clocks to be consistent, in agreement with the similarity in their ages that is reported in previous works. With this evidence we discarded some possible formation scenarios which could explain its chemical difference, like stellar migration or a chance alignment.

In order to attempt to test chemical clocks with a bigger and more representative sample of stars, we enlarged our sample by including 31 systems from the literature, finding that the components of these wide binaries follow a 1-1 relation in all the abundance ratios studied, as what is expected thanks to what we know about the formation scenarios of these systems. 
Our results indicate that systematic uncertainties from different methodologies can affect the final abundances, and probably have a significant effect on the different chemical clocks. In order to better assess and quantify this effect, a more careful analysis of common stars but different methods should be performed, as well as having the ages of all these stars. 

Also, we compared the degree of correlation of individual abundances and chemical clocks, and found that chemical clocks are more consistent between components of wide binaries than individual abundances, even for stars that might have gone through any process that alter its atmospheric composition. Therefore, we conclude that chemical clocks really carry more information than just the plain run of individual chemical abundances, information probably related to age and a common formation site. 

Finally, we propose that including them in the chemical tagging parameter space would be very beneficial and allow better identification of members of disrupted stellar groups.

\acknowledgments
This project is framed in the project “Gaia sibling stars as probes of the chemodynamical evolution of the Milky Way disk”, funded by ECOS-ANID grant number 180049. We acknowledge the anonymous referee for the valuable and useful comments that helped to improve the discussion of the paper. F.E-R. acknowledges Iván Ramírez, and appreciates the invaluable guidance and support of Marcelo Tucci-Maia, Claudia Aguilera and Danielle de Brito. P.J. acknowledges financial support from FONDECYT Iniciaci\'on Grant Number 11170174 and Regular grant Number 1200703. J.C. acknowledges support from CONICYT project Basal AFB-170002, and from the Agencia Nacional de Investigación y Desarrollo (ANID), via Proyecto FONDECYT Regular 1191366. LC acknowledge support from "Programme National de Physique Stellaire" (PNPS) and from the "Programme National Cosmologie et Galaxies" (PNCG) of CNRS/INSU.\\

\software{%BACCHUS code \citep{Masseron16}, 
iSpec \citep[][2019 version]{BlancoCuaresma14, BlancoCuaresma19}, MOOG \citep[][2017 version]{Sneden73, Sneden12}, and %, Qoyllur-Quipu \citep{Ramirez14}
IRAF \citep[][]{IRAF86,IRAF93}} %{\bf \color{blue} El referee sugiere incluir BACCHUS y q2, pero no estoy segura si debería ser el caso, ya que no los usé directamente}

\appendix

\section{Catalog}
\begin{deluxetable}{lll}[ht]
\tablenum{4}
\tablecaption{Summary of the whole sample of wide binaries used in this work. Astrometric data is taken from \textit{Gaia} DR3, except for radial velocities. The table is available as online material.}
\tablewidth{0pt}
\tablehead{
\colhead{Column} & \colhead{Units} & \colhead{Description}}
\startdata
        system     &         &   System name         \\
         Reference &         &   Paper reference where the data were taken from   \\
        Instrument &         &  Spectrograph used to obtained the data           \\
    Spectral range &      nm &  Spectral range of the instrument used           \\
              star &         &  Name of the star           \\
              Teff &       K &  Effective temperature          \\
             eTeff &       K &  Effective temperature error           \\
              logg &     dex &  Surface gravity           \\
             elogg &     dex &  Surface gravity error           \\
   $[\text{Fe/H}]$ &     dex &  Metallicity           \\
  e$[\text{Fe/H}]$ &     dex &  Metallicity error           \\
              vmic &    $\text{km s}^{-1}$ &  Microturbulence velocity           \\
             evmic &    $\text{km s}^{-1}$ &  Microturbulence velocity error           \\
                 s &      AU &  Projected physical separation of the two stars \\
    \hline
                ra &     deg &  Right ascension           \\
         ra\_error &     deg &  Right ascension error            \\
               dec &     deg &  Declination           \\
        dec\_error &     deg &  Declination error           \\
          parallax &     mas &  Parallax           \\
   parallax\_error &     mas &  Parallax error           \\
              pmra &  $\text{mas yr}^{-1}$ &  Proper motion in right ascension direction   \\
       pmra\_error &  $\text{mas yr}^{-1}$ &  Standard error of proper motion in right ascension direction           \\
             pmdec &  $\text{mas yr}^{-1}$ &  Proper motion in declination direction            \\
      pmdec\_error &  $\text{mas yr}^{-1}$ &  Standard error of proper motion in declination direction           \\
  dr2\_radial\_velocity &   $\text{km s}^{-1}$ &  Spectroscopic barycentric radial velocity from Gaia DR2           \\
  dr2\_radial\_velocity\_error &    $\text{km s}^{-1}$ & Standard error of spectroscopic barycentric radial velocity from Gaia DR2\\
phot\_g\_mean\_mag &     mag &  G-band mean magnitude (Vega scale)           \\
phot\_bp\_mean\_mag &     mag & Integrated BP mean magnitude            \\
phot\_rp\_mean\_mag &     mag & Integrated RP mean magnitude            \\
\enddata
\end{deluxetable}

\begin{longrotatetable}
\begin{deluxetable*}{lrrrrrrrrrrrrrr}
\tablenum{A1}
\tablecaption{Atomic data and equivalent widths of linelist used to derive stellar parameters}\label{table:stellarparam_atomicdata}
\tablewidth{0pt}
\tablehead{
\colhead{Element} & \colhead{$\lambda$} & \colhead{$\log gf$} & \colhead{$\chi$} & \colhead{Ref code} & \colhead{HIP15304} & \colhead{HIP15310} & \colhead{HIP32865} & \colhead{HIP32871} & \colhead{HIP34407} & \colhead{HIP34426} & \colhead{HIP52792} & \colhead{HIP52793} & \colhead{HIP58240} &  \colhead{HIP58241}\\
\colhead{} & \colhead{(\AA)} & \colhead{(dex)} & \colhead{(eV)} & \colhead{} & \colhead{} & \colhead{} & \colhead{} & \colhead{} & \colhead{} & \colhead{} & \colhead{} & \colhead{} & \colhead{} & \colhead{}}
\startdata
Fe 1 &  4802.880 & -1.514 &           3.642 &            BWL &     62.90 &     64.70 &     49.34 &     48.15 &     42.18 &     32.65 &     32.70 &     36.04 &     59.50 &     64.96 \\
Fe 1 &  4808.148 & -2.690 &           3.252 &            MRW &     30.41 &     31.84 &     18.46 &     16.86 &     13.85 &      8.93 &       NaN &     10.03 &     27.24 &     29.93 \\
Fe 1 &  4809.137 & -2.228 &           3.695 &            K07 &     29.30 &     29.65 &       NaN &     13.24 &       NaN &       NaN &       NaN &       NaN &       NaN &     31.01 \\
Fe 1 &  4909.383 & -1.231 &           3.929 &            K07 &     72.81 &       NaN &       NaN &       NaN &     43.15 &     33.04 &       NaN &       NaN &     68.42 &       NaN \\
Fe 1 &  4913.136 & -1.656 &           4.608 &            K07 &     21.51 &       NaN &       NaN &       NaN &      9.52 &       NaN &       NaN &       NaN &     19.19 &     23.09 \\
Fe 1 &  4917.230 & -1.080 &           4.191 &            MRW &     66.73 &     72.75 &     42.94 &     42.08 &     43.18 &     33.06 &       NaN &     36.92 &     68.24 &     78.59 \\
Fe 1 &  4962.572 & -1.182 &           4.178 &            BWL &     58.27 &       NaN &       NaN &       NaN &     30.31 &     21.55 &       NaN &       NaN &       NaN &       NaN \\
Fe 1 &  4986.223 & -1.290 &           4.218 &            MRW &     53.42 &     59.16 &     31.98 &     30.44 &     30.28 &     22.54 &     23.28 &     23.80 &     53.86 &     61.76 \\
Fe 2 &  4993.350 & -3.684 &           2.807 &             RU &     49.13 &     44.12 &     27.97 &     30.49 &     25.08 &     20.87 &     26.06 &     23.29 &       NaN &       NaN \\
Fe 1 &  4999.112 & -1.640 &           4.186 &            MRW &     34.30 &     37.47 &       NaN &       NaN &       NaN &       NaN &       NaN &       NaN &       NaN &       NaN \\
Fe 1 &  5023.186 & -1.500 &           4.283 &            MRW &     42.05 &     45.90 &     21.47 &     24.67 &     15.79 &     10.42 &     11.64 &     14.45 &     34.23 &     39.73 \\
Fe 1 &  5029.618 & -1.950 &           3.415 &            MRW &     53.93 &     56.83 &     37.44 &     34.77 &       NaN &       NaN &       NaN &       NaN &       NaN &     52.39 \\
Fe 1 &  5031.914 & -1.570 &           4.371 &            MRW &     30.82 &     34.77 &     15.42 &     15.22 &       NaN &       NaN &       NaN &       NaN &       NaN &       NaN \\
Fe 1 &  5036.922 & -3.010 &           3.017 &            MRW &     27.76 &     30.49 &     14.84 &     13.38 &       NaN &       NaN &       NaN &       NaN &       NaN &       NaN \\
Fe 1 &  5054.642 & -1.921 &           3.640 &            BWL &     46.23 &     48.99 &     30.04 &     28.78 &     20.70 &     13.82 &     14.07 &     17.34 &       NaN &     43.87 \\
Fe 2 &  5100.655 & -4.197 &           2.807 &             RU &     40.72 &       NaN &       NaN &     22.34 &     16.33 &     13.06 &     26.64 &     15.38 &       NaN &       NaN \\
Fe 1 &  5104.438 & -1.590 &           4.283 &            MRW &     42.67 &     46.28 &     23.81 &     23.48 &     20.25 &       NaN &       NaN &     15.59 &     35.86 &       NaN \\
Fe 1 &  5217.919 & -1.719 &           3.640 &            BWL &     55.28 &     58.03 &       NaN &     34.90 &       NaN &       NaN &     20.10 &       NaN &     50.64 &       NaN \\
Fe 1 &  5236.202 & -1.497 &           4.186 &            BWL &     43.27 &     44.81 &       NaN &       NaN &       NaN &       NaN &       NaN &       NaN &       NaN &     38.87 \\
Fe 1 &  5243.776 & -1.050 &           4.256 &            MRW &     72.79 &     75.25 &     52.22 &     51.01 &     45.20 &     35.81 &     38.63 &     39.69 &     66.54 &     74.68 \\
Fe 1 &  5253.021 & -3.840 &           2.279 &            MRW &     18.01 &     23.91 &     11.46 &      9.79 &      7.33 &       NaN &       NaN &       NaN &     18.58 &     22.93 \\
Fe 2 &  5264.802 & -3.130 &           3.230 &     2009A\&A... &     67.97 &       NaN &       NaN &     44.43 &     39.60 &     33.69 &     41.60 &     38.09 &     49.20 &     53.52 \\
Fe 1 &  5267.269 & -1.596 &           4.371 &             BK &     37.32 &       NaN &     19.14 &     16.90 &     11.01 &      6.76 &       NaN &      9.45 &     35.56 &       NaN \\
Fe 1 &  5288.525 & -1.493 &           3.695 &     BWL+2014MN &     67.40 &     69.82 &     46.62 &     43.24 &     38.21 &     29.01 &     33.26 &     32.38 &     59.56 &     66.78 \\
Fe 1 &  5293.959 & -1.770 &           4.143 &            MRW &     37.92 &     42.09 &     22.84 &     21.11 &     16.72 &     11.46 &     10.49 &     13.64 &     31.72 &     34.28 \\
Fe 1 &  5294.547 & -2.760 &           3.640 &            MRW &     16.13 &     19.96 &       NaN &       NaN &       NaN &       NaN &       NaN &       NaN &       NaN &       NaN \\
Fe 1 &  5295.312 & -1.590 &           4.415 &            MRW &     34.20 &     37.57 &     20.88 &       NaN &     15.56 &       NaN &     11.06 &     12.28 &     30.59 &     34.78 \\
 \enddata
\end{deluxetable*}
\movetabledown=20mm
\end{longrotatetable}

\begin{longrotatetable}
\begin{deluxetable*}{lrrrrrrrrrrrrrr}
\tablenum{6}
\tablecaption{ Atomic data and equivalent widths of linelist used to derive abundances}\label{table:abundances_atomicdata}
\tablewidth{0pt}
\tablehead{
\colhead{Element} & \colhead{$\lambda$} & \colhead{$\log gf$} & \colhead{$\chi$} & \colhead{Ref code} & \colhead{HIP15304} & \colhead{HIP15310} & \colhead{HIP32865} & \colhead{HIP32871} & \colhead{HIP34407} & \colhead{HIP34426} & \colhead{HIP52792} & \colhead{HIP52793} & \colhead{HIP58240} &  \colhead{HIP58241}\\
\colhead{} & \colhead{(\AA)} & \colhead{(dex)} & \colhead{(eV)} & \colhead{} & \colhead{} & \colhead{} & \colhead{} & \colhead{} & \colhead{} & \colhead{} & \colhead{} & \colhead{} & \colhead{} & \colhead{}}
\startdata
Cr 1 &  4801.025 & -0.131 &           3.122 &            MFW &     52.64 &     55.33 &     38.75 &     36.37 &     31.71 &     22.25 &       NaN &     24.84 &     49.73 &     54.27 \\
Fe 1 &  4802.880 & -1.514 &           3.642 &            BWL &     63.46 &     65.45 &     49.34 &     48.61 &     44.16 &     33.60 &     32.70 &     36.04 &     58.95 &     64.95 \\
Ti 2 &  4806.321 & -3.380 &           1.084 &            BHN &     14.29 &     12.90 &      9.35 &       NaN &      8.93 &       NaN &      7.31 &       NaN &     11.85 &       NaN \\   
Fe 1 &  4808.148 & -2.690 &           3.252 &            MRW &     30.83 &     32.31 &     19.78 &     16.86 &     14.98 &      9.60 &       NaN &     11.31 &     26.58 &     29.93 \\
Fe 1 &  4809.137 & -2.228 &           3.695 &            K07 &     29.60 &     29.97 &       NaN &       NaN &       NaN &       NaN &       NaN &       NaN &     25.03 &     28.86 \\
Zn 1 &  4810.528 & -0.160 &           4.078 &     1980A\&A... &     81.76 &     82.41 &     66.11 &     65.57 &     60.83 &     58.63 &     58.93 &     60.14 &     73.28 &     81.59 \\
Ni 1 &  4811.983 & -1.450 &           3.658 &     2003ApJ... &     27.69 &     28.82 &     17.06 &     16.05 &     12.32 &      8.80 &       NaN &      7.98 &     22.40 &     24.87 \\
Cr 2 &  4812.337 & -1.977 &           3.864 &            K10 &     52.72 &     50.87 &     30.60 &     34.24 &     28.89 &     22.36 &     23.56 &     23.63 &     42.93 &     47.29 \\
Co 1 &  4813.476 &  0.120 &           3.216 &            K08 &     48.96 &     52.93 &     34.09 &     30.11 &     24.82 &     18.09 &       NaN &     19.90 &     41.54 &     50.30 \\
Ni 1 &  4814.591 & -1.630 &           3.597 &     2014ApJS.. &     24.57 &     27.00 &     14.61 &       NaN &      9.91 &      6.71 &       NaN &       NaN &     18.38 &     21.87 \\
Zr 1 &  4815.630 & -0.030 &           0.604 &           BGHL &      8.08 &       NaN &       NaN &       NaN &       NaN &       NaN &       NaN &       NaN &       NaN &       NaN \\
Ti 1 &  4820.409 & -0.380 &           1.503 &     2013ApJS.. &     44.28 &     48.51 &     31.51 &     29.37 &     26.23 &     17.67 &       NaN &     18.24 &     42.41 &     47.13 \\
Mn 1 &  4823.520 &  0.136 &           2.319 &          DLSSC &    155.43 &    172.79 &    131.69 &    112.12 &    100.86 &     88.42 &     94.34 &    104.35 &    150.52 &    201.98 \\
Cr 2 &  4824.127 & -0.980 &           3.871 &            K10 &    103.69 &    103.37 &     74.37 &     81.33 &     74.01 &     62.51 &     77.50 &     71.27 &     98.03 &    110.96 \\
Zr 1 &  4828.040 & -0.640 &           0.623 &           BGHL &     10.46 &     11.19 &       NaN &       NaN &       NaN &       NaN &       NaN &       NaN &       NaN &       NaN \\
Cr 2 &  4836.229 & -1.960 &           3.858 &            SLd &     49.00 &     47.65 &       NaN &       NaN &       NaN &       NaN &       NaN &       NaN &       NaN &       NaN \\
Cr 2 &  4848.235 & -1.180 &           3.864 &            K10 &     34.77 &     31.06 &       NaN &       NaN &       NaN &       NaN &       NaN &       NaN &       NaN &       NaN \\
Co 1 &  4899.514 & -1.597 &           2.042 &            K08 &      5.41 &      6.65 &       NaN &       NaN &       NaN &       NaN &       NaN &       NaN &       NaN &       NaN \\
Y 2 &  4900.119 &  0.030 &           1.033 &          BBEHL &     85.77 &     93.15 &       NaN &       NaN &       NaN &       NaN &       NaN &       NaN &       NaN &       NaN \\\
Fe 1 &  4907.732 & -1.700 &           3.430 &       GESHRL14 &     60.37 &     65.31 &     28.16 &     25.81 &     42.75 &     31.51 &     28.06 &     32.58 &     71.50 &     89.12 \\
Fe 1 &  4908.599 & -4.160 &           2.484 &            FMW &      5.96 &       NaN &       NaN &       NaN &       NaN &       NaN &       NaN &       NaN &      6.34 &     19.71 \\
Fe 1 &  4910.325 & -0.459 &           4.191 &            K07 &    103.47 &    114.80 &     62.61 &     61.97 &     62.99 &     51.44 &       NaN &     50.39 &     89.34 &    172.72 \\
Ti 2 &  4911.194 & -0.640 &           3.124 &     2013ApJS.. &     73.88 &     74.88 &     38.34 &     43.81 &     45.65 &     39.48 &     46.97 &     39.75 &     57.31 &     59.45 \\
Fe 1 &  4913.136 & -1.656 &           4.608 &            K07 &     20.51 &       NaN &       NaN &       NaN &       NaN &      6.67 &       NaN &       NaN &     19.46 &     23.27 \\
Ti 1 &  4913.613 &  0.220 &           1.873 &     2013ApJS.. &     47.72 &     56.10 &     29.96 &     28.77 &     31.83 &     22.74 &     18.74 &     26.57 &     48.24 &     57.52 \\
Fe 1 &  4917.230 & -1.080 &           4.191 &            MRW &     66.42 &     71.83 &     42.94 &     42.08 &     46.41 &     35.26 &     32.78 &     36.92 &     73.99 &     78.59 \\
Fe 1 &  4918.012 & -1.260 &           4.231 &            MRW &     55.37 &     59.65 &     32.99 &     32.51 &     37.13 &     27.23 &     23.26 &       NaN &     57.31 &     55.75 \\
 \enddata
\end{deluxetable*}
\movetabledown=20mm
\end{longrotatetable}

\bibliography{bibliography}{}
\bibliographystyle{aasjournal}

\end{document}